\documentclass[dvips]{arxstspdf}
\usepackage{graphicx}
\usepackage{flushend}
\usepackage{stfloats}

\volume{23}
\issue{3}
\pubyear{2008}
\firstpage{287}
\lastpage{312}
\doi{10.1214/07-STS244}

\makeatletter
\setattribute{abstract}   {width}  {35pc}
\setattribute{keyword}    {width}  {35pc}

\newtheorem{lemma}{Lemma}[section]
\newtheorem{prop}{Proposition}[section]

\makeatother

\begin{document}
\begin{frontmatter}

\title{Quantifying the Fraction of Missing Information
for Hypothesis Testing in Statistical and Genetic Studies}
\runtitle{Relative Information}

\begin{aug}
\author[a]{\fnms{Dan L.} \snm{Nicolae}\corref{}\ead[label=e3]{nicolae@galton.uchicago.edu}},
\author[b]{\fnms{Xiao-Li} \snm{Meng}\ead[label=e2]{meng@stat.harvard.edu}} \and
\author[c]{\fnms{Augustine} \snm{Kong}\ead[label=e1]{kong@decode.is}}
\runauthor{D. L. Nicolae, X.-L. Meng and A. Kong}

\affiliation{The University of Chicago, Harvard University and deCode Genetics}

\address[a]{Dan L. Nicolae is Associate Professor, Departments of
Medicine and Statistics,
The University of Chicago, 5734 S. University Ave, Chicago, Illinois 60637,
USA \printead{e3}.}

\address[b]{Xiao-Li Meng is Whipple \mbox{V.~N.} Jones Professor and
Chair of Statistics, Harvard University, Cambridge, Massachusetts, USA
\printead{e2}.}

\address[c]{Augustine Kong is Vice President of Statistics,
deCode Genetics, Sturlugata 8, IS-101 Reykjavik, Iceland \printead{e1}.}

\end{aug}

\relateddoi{T2}{Discussed in \doi{10.1214/08-STS244B},
\doi{10.1214/08-STS244C}
and
\doi{10.1214/08-STS244A};
rejoinder at
\doi{10.1214/08-STS244REJ}.}

%
\begin{abstract}
Many practical studies rely on hypothesis testing procedures
applied to data sets with missing information. An important part of
the analysis is to determine the impact of the missing data on the
performance of the test, and this can be done by properly
quantifying the relative (to complete data) amount of available
information. The problem is directly motivated by applications to
studies, such as linkage analyses and haplotype-based association
projects, designed to identify genetic contributions to complex
diseases. In the genetic studies the relative information measures
are needed for the experimental design, technology comparison,
interpretation of the data, and for understanding the behavior of
some of the inference tools. The central difficulties in
constructing such information measures arise from the multiple,
and sometimes conflicting, aims in practice. For large samples, we
show that a satisfactory, likelihood-based general solution exists
by using appropriate forms of the relative Kullback--Leibler
information, and that the proposed measures are computationally
inexpensive given the maximized likelihoods with the observed
data. Two measures are introduced, under the null and alternative
hypothesis respectively. We exemplify the measures on data coming
from mapping studies on the inflammatory bowel disease and
diabetes. For small-sample problems, which appear rather
frequently in practice and sometimes in disguised forms (e.g.,
measuring individual contributions to a large study), the robust
Bayesian approach holds great promise, though the choice of a
general-purpose ``default prior'' is a very challenging problem. We
also report several intriguing connections encountered in our
investigation, such as the connection with the fundamental
identity for the EM algorithm, the connection with the second CR
(Chapman--Robbins) lower information bound, the connection with
entropy, and connections between likelihood ratios and Bayes
factors. We hope that these seemingly unrelated connections, as
well as our specific proposals, will stimulate a general
discussion and research in this theoretically fascinating and
practically needed area.
\end{abstract}

%
\pdfkeywords{EM algorithm, entropy, Fisher information, genetic linkage studies,
haplotype-based association studies, noninformative prior, Kullback--Leibler information,
relative information, Cox regression, partial likelihood}

\end{frontmatter}

\section{Many Challenges and An Overview}\label{sec:intro}

\subsection{General Challenges}\label{subsec:intro}

The central aim of this paper is to establish, in the context of
hypothesis testing with incomplete data, a~general framework for
quantifying the amount of information in the observed data for a
specific test being performed, relative to the full amount of
information we would have had the data been complete. We do not
address the issue of what is the best testing procedure, with or
without the complete data, nor the issue of whether a full
modeling/estimation strategy should or can be used instead.
Rather, we address an increasingly common practical problem where
the investigator has chosen the testing procedure, but needs to
know the impact of the missing data on the test in terms of the
relative loss of information. Such is the case in the genetic
studies we briefly review in Sections~\ref{sec:genet} and
\ref{sec:haplo}.

Besides the specific challenges listed in
Section~\ref{subsec:conf}, there are a number of general
theoretical and methodological difficulties for establishing this
general framework. First, unlike the similar task for estimation,
where the notion of ``fraction of missing information'' is well
studied and documented
(e.g., \cite*{demplairrubi77}; \cite*{mengrubi91}), for hypothesis
testing, there are two sets of measures to be contemplated,
depending on whether the null hypothesis or the posited
alternative model can be regarded as approximately adequate.
Indeed, this is the very question the hypothesis test aims to
provide partial evidence to discriminate.

Second, hypothesis testing procedures, especially those of
nonparametric or semiparametric nature, are often constructed
without reference to a specific (parametric) model. However,
without an explicit model to link the unobserved quantities with
the observed data, the very task of measuring how much information
we have missed is neither possible in general nor meaningful. It
is known, though not widely (e.g., \cite*{cher79}; \cite*{meng01}), that
certain robust statistical procedures for estimation or testing
can produce more efficient or powerful results with less data.
Consequently, without assuming that our testing procedure is
optimal under a specified optimality criterion, we may end up with
the seemingly paradoxical situation that additional data may make
our procedure less efficient or powerful. That is, we may declare
that more information is available with less data.

Third, in the context of small samples, quantifying information
requires going beyond convenient and standard measures such as
Fisher information, which is essentially a large-sample measure.
Small-sample problems are rather frequent with incomplete data, as
missing data reduce effective sample sizes. For the genetic
studies we investigate in this paper, the small-sample problems
arise even when there appear to be ample amounts of data. For
example, we are often interested in measuring information content
in individual components (e.g., an individual family in a large
linkage study). In haplotype association studies, we often test
haplotypes individually---data size may be large enough for
testing a common haplotype, but very small for a rare one. In
addition, an individual person can be fully informative for one
haplotype because we know s/he cannot carry it, but much less so
for another when we are uncertain whether s/he carries it or not.
All these problems remind us that, in general scientific studies,
small-sample problems appear more often than meets the eyes,
namely, the numerical value of the sample size, because they
sometimes appear in disguised forms.

Given the complex nature of small-sample problems requiring
information measures, we literally have spent several years in our
quest of finding a general workable approach. Not surprisingly,
our conclusion is that robust Bayesian methods hold more promise.
As we propose in Section~\ref{sec:small}, after establishing a
likelihood-based large-sample framework in
Section~\ref{sec:large}, this problem can be dealt with by
considering posterior measures of the flatness of the entire
likelihood surfaces. However, the problem of specifying an
appropriate ``default'' prior is challenging. We report both our
promising findings and open problems, hoping to stimulate further
development on this practically important and theoretically
fascinating topic. We also discuss various interesting theoretical
connections (Section~\ref{sec:theor}), as well as further
methodological work and applications (Section~\ref{sec:future}).

\subsection{Conflicting Aims in Genetic Studies}\label{subsec:conf}

The central applied problem that motivated our work was the task
to sensibly measure and efficiently compute the amount of
information available in \textit{a particular genetic data set for a
particular hypothesis tested by a particular statistical
procedure}. All genome-wide linkage screens carried out on
qualitative and quantitative traits as well as most of the
association studies extract only part of the underlying
information. Missing information can be the result of different
sources, such as absence of DNA samples, missing genotypes,
spacing between markers, noninformativeness of the markers, or
unknown haplotype phase. Investigators want to know how much
information is available in the observed data for the purpose of
the study \textit{relative} to the amount of information that would
have been available if the data were complete. The notion of
complete data is problem specific and, in parametric inference,
depends on the sufficient statistics; for example, in linkage
studies where the IBD (identical by descent) process is sufficient
for inference, complete data can be achieved even if genotypes
and/or individual samples are missing. Measures of relative
information are needed for designing follow-up strategies in
linkage studies, for example, using more genetic markers with existing
DNA samples versus collecting DNA samples from additional
families. Even for situations where we do not intend to recover
the missing data, including situations where they cannot possibly
be recovered (e.g., DNA samples from deceased ancestors), such
measures can still be useful for the interpretation of the data
and of the results, and for understanding the behavior of some of
the inference tools (e.g., see Section \ref{s4.5}).

The key methodological challenge is to find a measure that (1) is
a reliable index of the relative information specific to a study
purpose, (2) conditions on particular data sets, (3) is robust in
the sense of general applicability, including to small data sets,
(4) is easy to compute and (5) is subject to meaningful
combination axioms. The reliability criterion (1) is obvious, and
the criterion (2) is necessary because typically an investigator
is interested in measuring the relative information in the data
set at hand, not with respect to some ``average'' data set.
Criterion (3) is desirable because in a typical genetic linkage
study one needs to deal with a large amount of data with a variety
of different complex structures (e.g., from a nuclear family to a
very complex pedigree), often under time constraints, and thus it
is not feasible to design separate measures to suit particular
data structures. Criterion (4) is needed for similar reasons---any
method without suitable computational efficiency, regardless
of its theoretical superiority, will typically be ignored in
routine genetic studies given the practical constraints. Criterion
(5) ensures certain desirable coherence to prevent paradoxical
measure properties (e.g., more informative studies receive less
weight in the combined index) when combining studies.

To deal with all these criteria simultaneously requires a careful
combination of Bayesian and frequentist perspectives. Some of the
criteria [e.g., (1) and (2)] are most easily handled from the
Bayesian perspective, and some [e.g., (5)] are easier to satisfy
with a frequentist criterion. With large samples, as it is
typical, likelihood theory provides a rather satisfactory
solution, as we demonstrate in Section~\ref{sec:large}. For small
samples, we have not been able to find a better alternative than
to follow a robust Bayesian perspective, which takes full
advantage of the Bayesian formulation in deriving information
measures with desirable coherent properties, and at the same time
it seeks measures that are robust to various misspecifications and
are thus more generally applicable. We emphasize, however, that
the computational burden associated to these Bayesian measures
should not be overlooked, even in this age of the MCMC revolution,
for the reasons underlying criterion~(4) above. Nevertheless, it
is more principled and fruitful to seek ways to increase
computational efficiency \textit{after} we establish theoretically
sound measures. This is the route we follow.

\subsection{Imputing Under the Null or Not---Gaining~Insight}\label{subsec:insig}

For those who have no (direct) interest in genetic studies, the
following simple example may provide a stimulus to follow the
methods developed in our paper. The example also provides some
insights into a somewhat ``perplexing'' practical question when
dealing with hypothesis testing in the presence of missing data:
shall we impute under the null or not? We emphasize that the
purpose of this example is \emph{not} to illustrate imputation
methods. Indeed, neither method discussed below can be recommended
in general. Rather, it shows how we can quantify relative
information by measuring \textit{how inaccurate is} to erroneously
treat imputations as if they were observed data.

Specifically, suppose $y_1,\ldots,y_n$ are i.i.d. realizations of
Bernoulli$(\mathrm{p})$, but only $n_0 < n$ of them are actually observed.
Assuming that the missing data are missing completely at random
(\cite*{rubi76}), we can denote the observed data by
$y_1,\ldots,y_{n_0}$. Evidently, a simple large-sample test
(assuming $n_0$ is adequately large) for $H_0\dvtx p=p_0$ is to refer
the test statistic (where the subscript ``ob'' stands for
``observed data'')
%
\begin{equation}\label{toytest}
T_{\mathrm{ob}}=\frac{\bar y_{\mathrm{ob}}-p_0}{\sqrt{p_0(1-p_0)/n_0}}
\end{equation}
to the null distribution $N(0,1)$, where $\bar y_{\mathrm{ob}}$ is the
average of the observed data.

Let us assume that the missing $y$'s were imputed using two
mean-imputation methods. The first\break method is to impute each
missing $y$ by its mean, estimated by $\bar y_{\mathrm{ob}}$. The second
procedure is to impute each missing $y$ by its mean assuming
$H_0$ is true, that is, by $p_0$. Clearly, with either
imputation, if we treat the imputed data as if they were observed
and apply the test (\ref{toytest}) with $n_0=n$, we will not reach
the valid conclusion unless we adjust the null distribution
$N(0,1)$.

For the first method, the average of all data, observed and
imputed, is $\bar y^*_1=\bar y_{\mathrm{ob}}$. Therefore, if we erroneously
treat the imputed values as real observations, we would compute our
test statistics as
%
\begin{equation}\label{talte}
T_1^*=\frac{\bar y^*_1-p_0}{\sqrt{p_0(1-p_0)/n}}
= \frac{1}{\sqrt{r}}T_{\mathrm{ob}},
\end{equation}
where
$r=n_0/n$. In contrast, the second method would lead to
%
\begin{equation}\label{tnull}
T_0^*=\frac{\bar y^*_0-p_0}{\sqrt{p_0(1-p_0)/n}}
= \sqrt{r}T_{\mathrm{ob}},
\end{equation}
because the average of all data, observed and imputed, is
$\bar y^*_0=r\bar y_{\mathrm{ob}}+(1-r)p_0$.

Two aspects of the above calculations are important. First, in
both cases, the resulting ``completed-data'' test statistic is
proportional to the benchmark given in (\ref{toytest}).
Consequently, imputing under the null or not leads to the same
answer, as long as we adjust the corresponding null distribution
accordingly (the generality of this equivalence result obviously
needs qualification, but the validity of a test is automatic when
its null reference distribution is correctly specified). Second,
identities (\ref{talte}) and (\ref{tnull}) yield respectively
%
\begin{equation}\label{rform}
r= \biggl( \frac{T_{\mathrm{ob}}}{T_1^*}  \biggr)^2
\quad \mbox{and}\quad
r= \biggl( \frac{T_0^*}{T_{\mathrm{ob}}}\biggr)^2.
\end{equation}
The results in (\ref{rform}) are important because $r=n_0/n$
measures the relative sample sizes, and hence the ``relative
information'' in an i.i.d. setting. These results suggest that we
consider measuring the relative information by how liberal the
first imputation-based test is, when the imputations under the
alternative are treated as real data, or how conservative the
second test is, when the imputations under the null are treated as
real observations. Our general large-sample results given in
Section~\ref{sec:large} show that these ideas are in fact general,
once we replace the statistics in (\ref{rform}) by their
appropriate log-likelihood ratio counterparts (recall the
large-sample equivalence between log-likelihood ratio statistics
and the Wald statistics in a form similar to $T^2$). Readers who
are not interested in genetic applications can go directly to
Section~\ref{sec:large}, as Sections~\ref{sec:genet} and
\ref{sec:haplo} provide the necessary background on the
genetic problems to which our methods will be applied.

\section{Genetic Linkage Analysis}\label{sec:genet}

\subsection{Allele-Sharing Methods}\label{sec:gene}

\begin{figure}[b]

\includegraphics{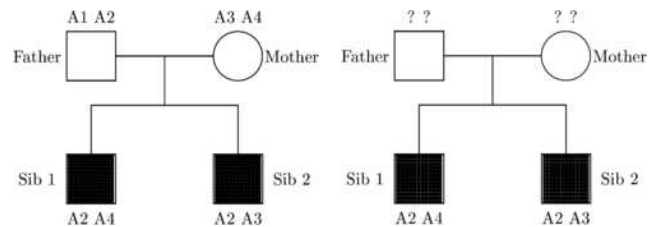}

\caption{Pedigree diagrams of an affected sib
pair; the IBD sharing is known for the sibs in the left diagram,
but only the IBS sharing is known for the sibs in the right
diagram.\label{fig:sibpair}}
\end{figure}

Linkage refers to the co-inheritance of two markers or genes
because they are located closely on the same chromosome.
Allele-sharing methods are part of linkage techniques for locating
regions on the genome that are very likely to contain disease
susceptibility genes (e.g., \cite*{ott91}). The data usually
consist of genotypes from a large number of markers (polymorphic
locations) spread along the genome for individuals from~$n$
pedigrees. The allele-sharing methods focus on affected
individuals, but genetic data on unaffected relatives are used to
infer the inheritance patterns. Alleles at the same locus in two
individuals are said to be identical by descent (IBD) if they
originate from the same chromosome, and are called identical by
state (IBS) if they appear to be the same. For a given location on
the genome, the evidence for a disease-susceptibility locus linked
to it is given by the sharing of alleles IBD among affected
relatives in excess of what is expected if the marker is not
linked to a genetic risk factor.

The simplest example of a data structure is the affected sib pair,
as shown in Figure \ref{fig:sibpair}, where the left diagram shows
a family with two affected brothers in which the parental
information at a fixed locus is denoted by ``A1'' and ``A2'' for
the father, and ``A3'' and ``A4'' for the mother. The siblings
have one allele IBD~(A2) which they inherited from their father,
and different alleles inherited from their mother. In general,
siblings share either two, one or no alleles IBD.
Unconditionally, each allele has probability $1/2$ to be transmitted;
this leads to a probability of $1/4$, $1/2$, $1/4$ for sharing zero,
one, two alleles, respectively, identical by descent. Conditioned
on the affection status of the sibs, in the neighborhood of a
disease gene, there is an expected increase in the number of
alleles IBD across a collection of sib pairs; statistical testing
methods are often used to measure the strength of the evidence.

In general, the data are not as simple as in the above example. The
pedigree structures can contain far more complicated relations
than sib pairs and more than two affected individuals. Most of the
data sets extract only part of the underlying IBD information. In
general, the information is incomplete at locations between
markers. Even at marker locations, a variety of factors can lead
to missing information, including any genotype data on deceased or
unavailable family members, missing genotypes in the typed
individuals, or noninformativeness of the markers. The right diagram
of \mbox{Figure}~\ref{fig:sibpair} illustrates a family where the parental allele
information is missing, so even though the allele sharing among the sib pair
appears to be identical in pattern with that of the left diagram,
it is not known if the sibs share one or zero alleles IBD as
the two ``A2'' alleles might originate on different parental chromosomes.

In general, the marker information of all the loci on the chromosome is
used to calculate a probability
distribution on the space of inheritance vectors. For locus~$t$
and pedigree $i$, an inheritance vector, $\omega_i=\omega_i(t)$,
is a binary vector that specifies, for all the nonfounding
members of the pedigree, which grand-parental alleles are
inherited. Under the assumption of no linkage, all inheritance
vectors are equally likely, which leads to a uniform prior
distribution on their space. For a sib pair, the inheritance
vector has four elements, one for each parent-child
combination. For example, the first element specifies whether the
allele inherited by the first sib from his father originates from
the grandfather or grandmother. Assuming no interference
(\cite*{ott91}), a Hidden
Markov Model can be used to calculate the inheritance distribution
conditional on the genotypes at all marker loci
(\cite*{landgree87}). The distribution of the inheritance vectors
conditional on the observed data is the basis of the statistical
inference, and it is used to determine the conditional
distribution of the number of alleles IBD at a given location.

\subsection{Hypothesis Testing Using Imputed Sharing~Scores}\label{subsec:share}

In order to summarize the evidence for
linkage in a pedigree, we can use a score $S_i$
(\cite*{whitha94}; \cite*{krug96}),
a measure of IBD sharing among the affected individuals at locus~$t$.
In general, $S_i$ is chosen such that it has a higher
expected value under linkage than under no linkage. The
standardized form of $S_i$ is 
$Z_i=(S_i-\mu_i)/\sigma_i$, where $\mu_i=\mathrm{E}(S_i|H_0)$ and
$\sigma_i^2 =\operatorname{Var}(S_i|H_0)$. The test is typically in
the form of linear combination over the $n$ pedigrees,
%
\begin{equation}\label{NPL}
Z=\frac{\sum_{i=1}^n \gamma_i Z_i}{\sqrt{\sum_{i=1}^n \gamma_i^2}},
\end{equation}
where $\gamma_i\geq0$ are weights assigned to the individual
families. The weights can be chosen according to the number of
affecteds and the relationship among them and/or other covariate
information. Under the null hypothesis, $Z$ has mean 0 and
variance 1. Deviations from the null hypothesis can be tested
using a $N(0,1)$ approximation or the exact distribution of $Z$.

In general, $Z_i$'s are not directly observable due to missing
information. A common practice is to impute/replace $Z_i$ by
$W_i=\mathrm{E}(Z_i|\hbox{data},H_0)$ to construct a test statistic
(\cite*{krug96}),
%
\begin{equation}\label{krugw}
W=\frac{\sum_{i=1}^n \gamma_i W_i}{\sqrt{\sum_{i=1}^n \gamma_i^2}}
= \mathrm{E}(Z|\hbox{data},H_0).
\end{equation}
The main problem with this test statistic is the difficulty of
directly evaluating its statistical significance. A standard
$N(0,1)$ approximation can be very inaccurate when there is a
large amount of missing information, as can be seen from the
following variance decomposition:
%
\begin{eqnarray}\label{vardec}
\operatorname{Var}(Z\vert H_0)
&=& \operatorname{Var}(\mathrm{E}(Z\vert\hbox{data}, H_0)\vert H_0)
\nonumber\\[-8pt]
\\[-8pt]
&&{} + \mathrm{E}(\operatorname{Var}(Z\vert\hbox{data}, H_0)\vert H_0),
\nonumber
\end{eqnarray}
which implies
%
\begin{equation}\label{varinq}
\quad
\operatorname{Var}(W\vert H_0)
= 1 - \mathrm{E}(\operatorname{Var}(Z\vert\hbox{data}, H_0)\vert H_0)\leq1.
\end{equation}
In many cases $\operatorname{Var}(W\vert H_0)$ can be substantially
less than 1,
leading to a conservative test when the $N(0,1)$ approximation is
used. A more accurate approach is described in Section
\ref{sec:exp}.

It is important to emphasize that, in allele-sharing studies, the
amount of missing information can be made arbitrarily low, at
least in theory, by increasing the number of markers in the
region. That is why, in regions with evidence for linkage, it is
important to predict whether by genotyping additional markers one
will obtain a more significant deviation from the null. A
different strategy for increasing the amount of information is to
increase the sample size, that is, to collect DNA samples from
additional families. Therefore knowing how much information is
missing from the data is important for designing efficient
follow-up strategies (see also \cite*{nico04}).

\subsection{Associating a Test With a Model}\label{sec:exp}

The linkage methods we described are based
on a chosen test statistic. In order to measure the relative
information for a test statistic, we need to associate it with a
model which specifies the stochastic relationship between the
observed data and missing data beyond the null. Otherwise the
question of relative information is not well defined, as it is
emphasized in \mbox{Section}~\ref{subsec:intro}. It has been shown by
\citet{kongcox97} that for every test statistic of the form of
(\ref{NPL}), a class of one-parameter models can be constructed
such that the efficient score (\cite*{coxhink74}) from each of
the models gives asymptotically equivalent results to the given
statistic. The inference procedures based on these models can be
applied to any pedigree structure and missing data patterns.

As an illustration, we briefly describe the \textit{exponential
tilting} model of \citet{kongcox97} applied to the one-locus
allele-sharing statistic. A key assumption underlying this model
(and other models for associating tests) is that the distribution
of the inheritance vectors satisfies
%
\begin{equation}\label{reduc}
\quad
\frac{P(\omega_i\vert H_A)}{P(\omega_i\vert H_0)}
= \frac{P(Z_i=z(\omega_i)\vert H_A)}{P(Z_i=z(\omega_i)\vert H_0)}
\quad \mbox{for all } i,
\end{equation}
where $\omega_i$ is an inheritance vector for pedigree $i$ that
leads to a standardized scoring function equal to $z(\omega_i)$,
and $H_A$ denotes the alternative hypothesis. Note that any time
an investigator employs a test solely based on the $Z$'s, as far as
measuring information concerns, s/he is effectively assuming
(\ref{reduc}) regardless of whether or not s/he is aware of it.

Under assumption (\ref{reduc}), it is sufficient to define
the alternative models for $Z_i$'s.
The exponential tilting model has the form
%
\begin{equation}\label{tilting}
P_\theta (Z_i=z)=P_0(Z_i=z) c_i(\theta ) \exp(\theta \gamma_iz),
\end{equation}
where $P_0(Z_i=z)$ is specified by the null (i.e., no linkage) and
$ c_i(\theta ) =[\sum_z P_0(Z_i=z)\exp(\theta \gamma_iz)]^{-1}$ is the
renormalization constant. When $Z$ is binary (e.g., as with
half-sibs), the model is the same as the logistic regression model
%
\begin{equation}\label{logit}
\operatorname{logit} P_\theta (Z_i=1) = \mu_i + \theta \gamma_i,
\end{equation}
where $\mu_i= {\rm logit} P_0(Z_i=1)$.

Given the exponential tilting model or other similar models (e.g.,
the linear model of \cite*{kongcox97}), the log-likelihood can be
calculated exactly for any missing data patterns under the
assumption (\ref{reduc}). Similar constructions can be done for
multilocus models, as in \citet{Nico99}.

\section{Haplotype-Based Association Studies}\label{sec:haplo}

\subsection{Basics of Association Studies}

Genetic association studies are designed to study potential
associations between genetic variants and phenotypes (i.e.,
observable traits) on a population scale. The association between
the genotype at a given marker and a disease can appear because
the genetic variant may be a risk factor for the disease, or
because the variant may be strongly correlated, called \textit{in
linkage disequilibrium} (LD) in the genetics literature, with a
causal locus. The magnitude of the correlation depends on many
factors including the distance between the markers and the
population history.

For the simplicity of description, we focus here on a simple and
popular design, case-control studies, although most results and
principles are applicable to other sampling designs including
those that incorporate quantitative traits and family-based
controls. The simplest genetic variant and a commonly used genetic
marker is a single nucleotide polymorphism (SNP) that takes on
only two possible alleles. Denoting the two possible alleles as 1
and 2, there are three possible genotypes $(1,1)$, $(1,2)$ and $(2,2)$.
The data for a case-control study can then be summarized as a 2-by-3
table where the entries are counts of the three genotype
categories for the cases and controls, respectively. The data can
be further reduced to a 2-by-2 table, where the entries are counts
of the alleles, if a multiplicative model
(\cite*{terw92}; \cite*{falk87}) for
allele-risk is
assumed. Note that \mbox{under} common assumptions, for a person randomly
selected from the population, the two alleles carried are in
Hardy--Weinberg equilibrium, that is, they are independent. This might
not be true for an affected individual if the genotypes
confer different risks, but it is true for the~multiplicative model.
Since this model is true under the null hypothesis
which assumes no difference between the two alleles, assuming the
multiplicative model for the purpose of testing does not affect
the validity of the p-values. Obviously the power could be reduced if
the specified model is different from the true alternative.

When the causal locus genotypes are not part of the data, or when
the LD between the markers is strong, it might be more efficient
to use more than one marker simultaneously. Most of these
multilocus approaches for fine-mapping of disease alleles are
based on haplotypes
(e.g., \cite*{mcpe99}; \cite*{pritchard00}; \cite*{lam00};
\cite*{morr02}; \cite*{pritchard05}).
Haplotype analyses can be used to investigate untyped genetic
variation (\cite*{peer06}; \cite*{nicolae06}), and can be used to explore which
markers could be causal and which are unlikely to be so.
A \textit{haplotype} is a sequence of alleles along a chromosome, and hence
each person has two haplotypes. The alleles appearing in a
haplotype are said to be in \textit{phase}. If the haplotypes are
directly observed, then standard methods for analyzing contingency
tables could be used to test various models (\cite*{gret03}).
Possible scenarios range from having a candidate at-risk haplotype
to testing the full model (all the haplotypes have different
risks) versus the null model (all the haplotypes have the same
risk).

\subsection{Causes of Incomplete Information}

With a case-control study conducted with individual SNPs
separately, the sufficient statistic is a 2-by-2 table under the
multiplicative model and a likelihood ratio $\chi^2$ test can be
used to test the null hypothesis.
A common cause of incomplete information
is missing genotypes since yield is often less than perfect.
The situation becomes more
complicated when multiple SNPs are considered jointly. With two SNPs,
both having alleles denoted with 1 and 2, there are four possible
haplotypes: 1-1 (characterized by allele 1 at both SNPs), 1-2, 2-1 and
2-2. One
simple alternative hypothesis is that haplotype 1-1 has risk that
is different from the other three haplotypes which are assumed to
have the same risk. It could be that we believe the two SNPs are
functional and there is interaction between them that leads to
increased disease risk for haplotype 1-1, but more common is the
hypothesis that
the putative, but unobserved, mutation occurred in the 1-1
background and the association between the haplotype and the trait
is a result of both being associated with the mutation.

Under the multiplicative model, if haplotypes can be observed
directly, then this problem can again be reduced to a 2-by-2 table
of haplotype counts where the haplotypes 1-2, 2-1 and 2-2 are
collapsed into one. However, for the commonly used technology,
SNPs are genotyped separately.
For an individual, apart from incomplete information due to missing
the genotype for one of the SNPs, there is the issue of uncertainty in
phase. Specifically, if the genotypes for
the first and second SNP are $(1,2)$ and $(1,2)$ respectively, then
the two haplotypes could be either (1-$1,2$-2) or (1-$2,2$-1), that is,
the information on phase is missing. In general, there is
incomplete information on phase if two or more SNPs that make up
the haplotype are heterozygous.
In family-based association studies
(e.g., \cite*{abecasis00}; \cite*{martin00};
\cite*{langelaird02a}, \citeyear{langelaird02b}),
the data on relatives will provide additional
information on phase but
there will still be uncertainty in inferring the haplotypes.
For SNPs that are close together
physically, there exist typing technologies that can determine
the haplotypes directly, but they are usually much more expensive.
Hence, from the design perspective, quantifying loss of
information is relevant not only for power/sample-size
calculations, but also for the choice of technology.

%
\subsection{Measuring Relative Information Via Test
Statistics---a Two-Sample Example}\label{subsec:hapo}

Apart from being relevant for experimental design and the
interpretation of data, the amount of missing information is also
useful for understanding the behavior of certain testing
procedures. While one obvious way to perform testing is to apply a
likelihood ratio test based on actual likelihoods computed for the
observed incomplete data under the null hypothesis and alternative
hypothesis separately, software for such calculations which allows
the user to define models in a flexible manner is not readily
available. However, available are methods and software based on
the EM algorithm that can be applied to one sample to calculate
maximum likelihood estimates of haplotype frequencies and expected
haplotype counts for individuals or groups assuming the maximum
likelihood estimates are the true parameter values
(\cite*{exc95}; \cite*{haw95}; \cite*{long95}).
Other more sophisticated methods
and software to predict haplotype phase and estimate counts also
exist (e.g., \cite*{step01}; \cite*{niu02}). It is very tempting
for the user to apply standard testing
procedures, such as the likelihood ratio test, by simply treating
these expected/predicted counts as the actual observed counts.
Doing this is analogous to the example in
Section~\ref{subsec:insig}, except here we are dealing with a
two-sample problem.

Specifically, if the original EM computation is applied to the
cases and controls jointly as a single group (i.e., as under the
null), but with the expectation counts tabulated for the
individuals who are then separated into cases and controls, the
test is conservative. If, however, the EM computation is applied
to the cases and controls separately, then the result is
anti-conservative. Moreover, the degree of conservativeness with
the first procedure, in large samples, matches the degree of
anti-conservativeness of the second procedure. To be more
specific, consider the following simple example. Suppose the
observed data consist of 250 patients and 250 controls, or 500
chromosomes each. For a SNP, the patient counts are 300 allele 1
and 200 allele 2, and the control counts are 250 allele 1 and 250
allele 2. Let $a$ and $u$ denote respectively the population
frequency of allele 1 in cases and controls. Under the null, the
maximum likelihood estimates are $\tilde{a} = \tilde{u} =
(300+250)/(500+500) = 0.55$ and the maximum likelihood estimates
under the alternative are $\hat{a} = 300/500 = 0.6$ and $\hat{u} =
250/500 = 0.5$. Simple calculations show that the log-likelihood
ratio $\chi^2$ statistic is
\[
2 [\ell(\hat{a},\hat{u})-\ell(\tilde{a},\tilde{u})] = 10.12.
\]
Now suppose there are another 250 cases and 250 controls each with
no data yet. Suppose we treat these as missing data and apply the
EM computation to the cases and controls jointly. Since $\tilde{a}
= \tilde{u} = 0.55$, these extra cases and controls each have
expected counts of 275 allele 1 and 225 allele 2. Together with
the original counts, this gives 575 allele 1 and 425 allele 2 for
the cases, and 525 allele 1 and 475 allele 2 for the controls. The
log-likelihood ratio $\chi^2$ statistic computed based on these
counts is 5.05, approximately one-half of 10.12.

By contrast, suppose the expected counts for the missing data are
computed for the cases and controls separately. In this case, the
presumed counts are simply twice the original counts: 600 allele 1
and 400 allele 2 for the cases, and 500 allele 1 and 500 allele 2
for the controls. The log-likelihood ratio $\chi^2$ statistic
computed from these counts is 20.24, or exactly double that of
10.12. While this example is extremely simple and unrealistic, the
phenomenon seen does extend to real data with haplotypes. Indeed,
this is just another example of the relationships given in
(\ref{rform}). That is, either ratio will correctly estimate that
the relative information is about 50\%. The theoretical results in
the next section provide a general framework for such estimation.

\section{A Large-Sample Framework}\label{sec:large}

\subsection{Variations on the EM Identity}

Our large-sample framework is built upon a simple identity
involving expected log-likelihood ratios, where the expectation is
with respect to the conditional distribution of the missing data
given the observed data. Expected lod scores have also been used
in the genetics literature to measure the information content of
the data (\cite*{ott2001}), and to investigate optimality and
validity of analytic strategies (e.g., \cite*{elston98};
\cite*{abreu99}; \cite*{daw00}). Note that lod stands for logarithm (usually
base 10) of the odds, and is used as a statistic for testing
whether two loci are linked.

Specifically, let $Y_{\mathrm{co}}$ be the complete data and
$Y_{\mathrm{ob}}$ be the observed data---note that
here $Y_{\mathrm{ob}}$ is a function of $Y_{\mathrm{co}}$. Let
$\ell(\theta |D)$ be the
log-likelihood of $\theta $ given data~$D$. Then for any $\theta _1$ and
$\theta _2$,
%
\begin{eqnarray}\label{decom}
&& \ell(\theta _1|Y_{\mathrm{co}}) - \ell(\theta_2|Y_{\mathrm{co}})
\nonumber\\
&&\quad = [\ell(\theta _1|Y_{\mathrm{ob}})
- \ell(\theta _2|Y_{\mathrm{ob}})]
\nonumber\\[-8pt]
\\[-8pt]
&&\qquad {}
+ [\log f(Y_{\mathrm{co}}|Y_{\mathrm{ob}}, \theta_1)
\nonumber \\
&&\hspace*{37pt}{} -
\log f(Y_{\mathrm{co}}|Y_{\mathrm{ob}}, \theta _2)].
\nonumber
\end{eqnarray}
By taking conditional expectation with respect to
$f(Y_{\mathrm{co}}|Y_{\mathrm{ob}}, \theta )$,
where $\theta $ is to be chosen, we have
%
\begin{eqnarray}\label{key}
&& \mathrm{E} [\mathrm{lod}(\theta _1, \theta _2|
Y_{\mathrm{co}})|Y_{\mathrm{ob}}, \theta ]
\nonumber\\
&&\quad = \mathrm{lod}(\theta _1, \theta _2|Y_{\mathrm{ob}})
\\
&&\qquad {} + \mathrm{E}
\biggl[\log{f(Y_{\mathrm{co}}|Y_{\mathrm{ob}}, \theta _1)\over
f(Y_{\mathrm{co}}|Y_{\mathrm{ob}},
\theta _2)}  \bigg|Y_{\mathrm{ob}}, \theta \biggr],
\nonumber
\end{eqnarray}
where $\mathrm{lod}(\theta _1, \theta _2|D)$ is the log of odds of
$\theta _1$ over
$\theta _2$ given data $D$. Here $\log$ can be of any base, and lod is
the log of the likelihood ratio, or more generally the log of
posterior ratios. Identity (\ref{key}) is a simple extension of
the key identity given in \citet{demplairrubi77} for the EM
algorithm. Specifically, using the notation of
Dempster, Laird and Rubin (\citeyear{demplairrubi77}) 
%
\begin{eqnarray}\label{eq:em}
Q(\theta |\theta ')
&=& \mathrm{E}[\ell(\theta |Y_{\mathrm{co}})|Y_{\mathrm{ob}},\theta ']
\quad \mbox{and} \quad
\nonumber\\[-8pt]
\\[-8pt]
H(\theta |\theta ')&=&\mathrm{E}[\log f(Y_{\mathrm{co}}|
Y_{\mathrm{ob}}, \theta )|Y_{\mathrm{ob}}, \theta '],
\nonumber
\end{eqnarray}
identity (\ref{key}) is the same as
%
\begin{eqnarray}\label{eq:emex}
\quad
&& Q(\theta _1|\theta )-Q(\theta _2|\theta )
\nonumber\\[-8pt]
\\[-8pt]
&&\quad = \ell_{\mathrm{ob}}(\theta _1) - \ell_{\mathrm{ob}}(\theta _2)
+ H(\theta _1|\theta )-H(\theta _2|\theta ),
\nonumber
\end{eqnarray}
where $\ell_{\mathrm{ob}}(\theta )\equiv\ell(\theta |Y_{\mathrm{ob}})$. In
Dempster, Laird and Rubin (\citeyear{demplairrubi77}), (\ref{eq:emex}) was given with
$\theta =\theta _2$, and was the basis for establishing the celebrated
monotone convergence property of the EM algorithm. As we shall
see, this intrinsic connection with the EM algorithm not only
helps greatly our theoretical development in
Section~\ref{sec:theor}, but more importantly it enables us to
compute our information measures directly from quantities that are
already used for the EM computation.

Intuitively, if $\theta _1$ is the truth, then if we had more data, which
would come from
$f(Y_{\mathrm{co}}|Y_{\mathrm{ob}}, \theta _1)$,
we would on average have a larger lod score
than $\mathrm{lod}(\theta _1, \theta _2|Y_{\mathrm{ob}})$. Indeed,
by taking $\theta =\theta _1$ in (\ref{key}) we see
%
\begin{eqnarray}\label{inealt}
&& \mathrm{E} [\mathrm{lod}(\theta _1,\theta_2|Y_{\mathrm{co}})
|Y_{\mathrm{ob}}, \theta _1 ]
\nonumber \\
&&\quad
= \mathrm{lod}(\theta _1, \theta_2|Y_{\mathrm{ob}})
+ \mathrm{KL}(\theta _1\dvtx \theta _2)
\\
&&\quad \ge
\mathrm{lod}(\theta _1, \theta _2|Y_{\mathrm{ob}}),
\nonumber
\end{eqnarray}
where $\mathrm{KL}(\theta _1\dvtx \theta _2)\ge0$ is the
Kullback--Leibler informa\-tion---in
favor of $\theta _1$ against $\theta _2$ when $\theta _1$ is
true---contained in the conditional distribution of $Y_{\mathrm{co}}$ given
$Y_{\mathrm{ob}}$.
The inequality in (\ref{inealt}) becomes equality if and only if
$\mathrm{KL}(\theta _1\dvtx \theta _2)=0$, which happens if and only if
$f(Y_{\mathrm{co}}|Y_{\mathrm{ob}},
\theta _1) = f(Y_{\mathrm{co}}|Y_{\mathrm{ob}}, \theta _2) $ (a.s.);
that is, given $Y_{\mathrm{ob}}$, the
additional data would contain no information to discriminate
$\theta _2$ from $\theta _1$. The Kullback--Leibler distance has been used
extensively in information theory (e.g., \cite*{cover91}) and
mathematical statistics (e.g., \cite*{aitchison75}). Recent work
on using K--L loss includes \citet{george05} and references therein.

Similarly, if $\theta _2$ is the truth, then on average we would
expect a
smaller $\mathrm{lod}(\theta _1, \theta _2|Y_{\mathrm{co}})$ if we
had observed $Y_{\mathrm{co}}$.
Mathematically, this is shown by taking
$\theta =\theta _2$ in (\ref{key}), which leads to
%
\begin{eqnarray}\label{inenul}
&& \mathrm{E} [\mathrm{lod}(\theta _1,
\theta _2|Y_{\mathrm{co}})|Y_{\mathrm{ob}}, \theta _2 ]
\nonumber \\
&&\quad = \mathrm{lod}(\theta _1, \theta _2|Y_{\mathrm{ob}})
- \mathrm{KL}(\theta _2\dvtx \theta _1 )
\\
&&\quad
\le\mathrm{lod}(\theta _1,\theta _2|Y_{\mathrm{ob}}),
\nonumber
\end{eqnarray}
and the inequality becomes equality if and only if, as before,
$f(Y_{\mathrm{co}}|Y_{\mathrm{ob}}, \theta _1)
= f(Y_{\mathrm{co}}|Y_{\mathrm{ob}}, \theta _2) $.

It is important to emphasize that because all the expectations
above are conditional upon $Y_{\mathrm{ob}}$, it is legitimate to
allow any of
the $\theta $'s to depend on $Y_{\mathrm{ob}}$. In particular, the
null value $\theta _0$ in the rest of this paper can be either a known fixed
value when $H_0$ is a sharp null, or more generally the
constrained MLE of~$\theta $ from $\ell(\theta |Y_{\mathrm{ob}})$
under the null. It
is also important to emphasize that although in this section we
focus on large-sample measures primarily because of their reliance
on maximum likelihood estimators (MLEs), as discussed below, all
the equalities and inequalities discussed above do not involve any
approximation, large sample or not. Therefore all measures
discussed below can also be very useful for small samples, as long
as the MLEs can be trusted (e.g., a small-sample MLE can have good
properties, such as under the normal models).

\subsection{A Large-Sample Measure of Relative Information
Against $H_0$}

Suppose the null value is $\theta _0$ and that the MLE of $\theta $
(under $H_1$) given $Y_{\mathrm{ob}}$ is $\theta _{\mathrm{ob}}$,
and $\mathrm{lod}(\theta _{\mathrm{ob}}, \theta _0|Y_{\mathrm{ob}})
(\ge0)$ is used to assess the evidence against
$H_0\dvtx \theta =\theta _0$. To avoid technical complexity that is not of general
interest for our proposals, we will assume (I)
$\theta _{\mathrm{ob}}$ is unique,
an assumption typically automatic with large samples, and (II)
$\theta _{\mathrm{ob}}\not=\theta _0$, an assumption rarely, if
ever, violated in
practice. (Nevertheless, for theoretical completeness, we will
consider the case of $\theta _{\mathrm{ob}}=\theta _0$ in
Section~\ref{sec:theor} via
a limiting argument.) Then, if we intend to measure the
information in the unobserved data for discrediting $H_0$, under
the large-sample assumption, a~natural thing to do is to treat
$\theta _{\mathrm{ob}}$ as the ``truth,'' and measure the expected
loss of $\mathrm{lod}$
in favor of $\theta _{\mathrm{ob}}$ relative to the expected
complete-data $\mathrm{lod}$
score. Namely, we can naturally define
%
\begin{eqnarray}\label{fracoa}
\mathcal{R}I_1 &=&{\mathrm{lod}(\theta _{\mathrm{ob}},
\theta _0|Y_{\mathrm{ob}})\over\mathrm{E}
[\mathrm{lod}(\theta _{\mathrm{ob}}, \theta_0|Y_{\mathrm{co}})
|Y_{\mathrm{ob}},\theta _{\mathrm{ob}} ]}
\nonumber\\[-8pt]
\\[-8pt]
&=& {\ell_{\mathrm{ob}}(\theta _{\mathrm{ob}}) -\ell_{\mathrm{ob}}
(\theta _0) \over Q(\theta _{\mathrm{ob}}|\theta _{\mathrm{ob}})
- Q(\theta _0|\theta _{\mathrm{ob}})}.
\nonumber
\end{eqnarray}
The last expression shows that the computation of $\mathcal{R}I_1$ only
requires evaluations, at $\theta =\theta _0$ and $\theta =\theta
_{\mathrm{ob}}$, of the
observed-data log-likelihood $\ell_{\mathrm{ob}}(\theta )$ and the
$Q$ function,
where the latter is readily available from the EM algorithm.

Under assumptions (I) and (II), $\mathcal{R}I_1$ is well\break defined and by
(\ref{inealt}), $0<\mathcal{R}I_1\le1$. It is 1 if and only if
$\mathrm{KL}(\theta _{\mathrm{ob}}\dvtx \theta _0)=0$, which means that
the missing data\break cannot
distinguish between $\theta _{\mathrm{ob}}$ and $\theta _0$ and thus
there is no
missing information given $Y_{\mathrm{ob}}$. It approaches 0 if and
only if $\mathrm{lod}(\theta _{\mathrm{ob}}, \theta _0|Y_{\mathrm{ob}})/
\mathrm{KL}(\theta _{\mathrm{ob}}\dvtx
\theta _0)\rightarrow0$, which makes sense because
if the observed-data likelihood has
diminishing ability, relative to that of the missing-data model
[as measured by $\mathrm{KL}(\theta _{\mathrm{ob}}\dvtx\theta _0)$], to
distinguish between
$\theta _{\mathrm{ob}}$ and $\theta _0$, then as far as providing
evidence \textit{against} $H_0$, the missing information approaches 100\%. One
very appealing feature of $\mathcal{R}I_1$ is its direct interpretability.
As seen in the haplotype example in Section~\ref{subsec:hapo}, a
value of $\mathcal{R}I_1=0.5$ implies that if we had the complete data,
the lod score would be expected to be twice $(\mathcal{R}I_1^{-1}=2)$ as
large.

When $\ell(\theta |Y_{\mathrm{co}})$ is linear in a
(multidimensional) summary
statistics (i.e., a complete-data sufficient
statistics) $S(Y_{\mathrm{co}})$,
as when the complete-data model is from an exponential family,
$\mathrm{lod}(\theta _{\mathrm{ob}}, \theta _0|Y_{\mathrm{co}})$
can be written as $\mathrm{lod}(\theta _{\mathrm{ob}},
\theta _0|S(Y_{\mathrm{co}}))$ and
\[
\mathrm{E} [\mathrm{lod}(\theta _{\mathrm{ob}}, \theta_0|
Y_{\mathrm{co}}) |Y_{\mathrm{ob}}, \theta _{\mathrm{ob}}]
= \mathrm{lod}(\theta _{\mathrm{ob}},\theta _0|S^*(Y_{\mathrm{ob}})),
\]
where $S^*(Y_{\mathrm{ob}})= \mathrm{E}[S(Y_{\mathrm{co}})
|Y_{\mathrm{ob}}, \theta _{\mathrm{ob}}]$. That is,
$\mathcal{R}I_1$ measures the anti-conservativeness of the completed-data
test by pretending that the actual value of the unobserved
$S(Y_{\mathrm{co}})$ is the same as its imputation under the (estimated)
alternative. Therefore, $\mathcal{R}I_1$ is the general version of the
first case in (\ref{rform}).

This measure also has the following property when combining data
sets. Suppose
$Y_{\mathrm{co}}=\{Y_{\mathrm{co}}^{(1)}, \ldots,Y_{\mathrm{co}}^{(n)}\}$ are mutually
independent and we define $\mathcal{R}I_i$ for each
$Y_{\mathrm{co}}^{(i)}$ as in (\ref{fracoa})
but using $\theta _{\mathrm{ob}}$ instead of individual
$\theta _{\mathrm{ob}}^{(i)}$ (i.e., an MLE based on
$Y_{\mathrm{ob}}^{(i)})$; then the overall
$\mathcal{R}I$ is a weighted harmonic mean of $\mathcal{R}I_i$'s weighted by
the individual lod score,
$\mathrm{lod}_i = \mathrm{lod}(\theta_{\mathrm{ob}},\theta _0|Y_{\mathrm{ob}}^{(i)})$,
namely,
%
\begin{equation}\label{eq:comb}
\mathcal{R}I_1^{-1} ={\sum^n_{i=1}\mathrm{lod}_i
\mathcal{R}I_{1,i}^{-1} \over\sum^n_{i=1}\mathrm{lod}_i}.
\end{equation}
However, the individual lod score, $\mathrm{lod}_i$, is not necessarily
always positive in practice, a problem that is closely related to
the problem of defining relative measures for small data sets
(e.g., for individual family), as discussed in
Section~\ref{sec:small}. Note that $\mathcal{R}I_1$ can also be expressed
as weighted arithmetic mean of $\mathcal{R}I_{1,i}$ if we choose the
weights to be proportional to the expected individual
complete-data lod score
$\mathrm{lod}_i^{(c)}=E[\mathrm{lod}(\theta _{\mathrm{ob}},\theta
_0|Y_{\mathrm{co}}^{(i)})|Y_{\mathrm{ob}}^{(i)}]$:
%
\begin{equation}\label{eq:combl}
\mathcal{R}I_1={\sum^n_{i=1}\mathrm{lod}_i^{(c)}
\mathcal{R}I_{1,i} \over\sum^n_{i=1}\mathrm{lod}_i^{(c)}}.
\end{equation}
Clearly (\ref{eq:comb}) and (\ref{eq:combl}) are equivalent, as
long as\break  \mbox{$\mathcal{R}I_{1,i}>0$}. The harmonic rule (\ref{eq:comb}) is
somewhat more appealing because of the direct interpretation of
the weight $\mathrm{lod}_i$.

\subsection{A Large-Sample Measure of Relative Information
Under $H_0$}

Inequality (\ref{inenul}) also suggests a large-sample measure of
the relative information under $H_0$. By taking $\theta _1=\theta $
and $\theta_2=\theta _0$ in (\ref{inenul}) we obtain that
%
\begin{eqnarray}\label{revi}
&& \mathrm{E} [\mathrm{lod}(\theta ,\theta _0|Y_{\mathrm{co}})
|Y_{\mathrm{ob}}, \theta _0 ]
\nonumber \\
&&\quad = \mathrm{lod}(\theta , \theta_0|Y_{\mathrm{ob}})
- \mathrm{KL}(\theta _0\dvtx \theta )
\\
&&\quad
\le \mathrm{lod}(\theta , \theta _0|Y_{\mathrm{ob}}).
\nonumber
\end{eqnarray}
Thus, when the additional data are from
$f(Y_{\mathrm{co}}|Y_{\mathrm{ob}}, \theta _0)$,
the expected complete lod score cannot exceed
the one based on the observed data, for any $\theta $. We can use
$\max_{\theta } \mathrm{E} [\mathrm{lod}(\theta ,
\theta _0|Y_{\mathrm{co}}) |Y_{\mathrm{ob}}, \theta _0]$,
which cannot exceed $\mathrm{lod}(\theta _{\mathrm{ob}}, \theta_0|Y_{\mathrm{ob}})$ by
(\ref{revi}), as our best estimate of the complete-data lod score; the
use of a single point estimate of the complete-data
lod score without considering its
uncertainty can be justified under the large-sample assumption.
Consequently, we can define
%
\begin{eqnarray}\label{fracmb}
\mathcal{R}I_0 &=& {\max_{\theta }\mathrm{E} [\mathrm{lod}(\theta ,
\theta _0|Y_{\mathrm{co}})|Y_{\mathrm{ob}}, \theta _0 ]
\over\mathrm{lod}(\theta _{\mathrm{ob}}, \theta _0|Y_{\mathrm{ob}})}
\nonumber\\[-8pt]
\\[-8pt]
&=& \frac{\max_{\theta }[Q(\theta |\theta _0) - Q(\theta _0|
\theta_0)]}{\ell_{\mathrm{ob}}
(\theta _{\mathrm{ob}})-\ell_{\mathrm{ob}}(\theta _0)}.
\nonumber
\end{eqnarray}
The last expression shows again the computational efficiency of
this measure because $\max_{\theta }Q(\theta |\theta _0)$ is the
same as carrying out the E-step and M-step of an EM algorithm, by
pretending the previous iterated value is $\theta =\theta _0$. However,
we emphasize that the use of
$\max_{\theta } \mathrm{E} [\mathrm{lod}(\theta ,
\theta _0|Y_{\mathrm{co}}) |Y_{\mathrm{ob}}, \theta _0 ]$ in our
definition of $\mathcal{R}I_0$
instead of $\mathrm{E} [\max_{\theta } \mathrm{lod}(\theta ,
\theta _0|Y_{\mathrm{co}}) |Y_{\mathrm{ob}},
\theta _0 ]$ is not because this computation is easy, but rather
because of the nature of the fundamental identity (\ref{key}),
which requires we maximize the expected complete-data lod score.

Like $\mathcal{R}I_1$, $0 \le\mathcal{R}I_0\le1$. Unlike $\mathcal{R}I_1$, however,
the investigation of when $\mathcal{R}I_0$ approaches one or zero is a
more complicated matter, especially when the difference between
$\theta _{\mathrm{ob}}$ and $\theta _0$ is large. This is a partial
reflection of the
fact that $\mathcal{R}I_0$ is defined under the assumption that the null
hypothesis is (approximately) valid, which would be contradicted
by a large value of $\delta=\theta _{\mathrm{ob}}- \theta _0$,
especially under our
large-sample assumption. Therefore, it is more sensible to
investigate its theoretical properties when $\delta$ is small, in
which case it is essentially equivalent to $\mathcal{R}I_1$, as we will
establish in Section~\ref{sec:theor}. Nevertheless, it is useful to remark here
that under the additional assumption that $\theta _{\mathrm{ob}}$ is
the unique
stationary point of $\ell_{\mathrm{ob}}(\theta )$, the numerator of
$\mathcal{R}I_0$ is
zero if and only if its denominator is zero, that is, if and only
if $\ell_{\mathrm{ob}}(\theta _{\mathrm{ob}})=\ell_{\mathrm{ob}}(\theta _0)$.
[The ``if'' part of this result
is a trivial consequence of (\ref{revi}). The ``only if'' part
follows from the fact that if the numerator is zero, then $\theta _0$
is a maximizer of $Q(\theta |\theta _0)$, which means that $\theta_0$ must
also be a stationary point of $\ell_{\mathrm{ob}}(\theta )$ by
(\ref{eq:meng}) in Appendix~\ref{sA.2}.] This demonstrates
that in order for $\mathcal{R}I_0$ to
be very small, the observed-data likelihood must suffer a
diminishing ability to distinguish between $\theta _{\mathrm{ob}}$
and $\theta _0$, just as with $\mathcal{R}I_1$.

Also as with $\mathcal{R}I_1$, when $\ell(\theta |Y_{\mathrm{co}})$
is linear in $S(Y_{\mathrm{co}})$,
$\mathcal{R}I_0$ can be computed simply as
\[
\mathcal{R}I_0= {\max_\theta \mathrm{lod}(\theta , \theta_0|S^*_0(Y_{\mathrm{ob}}))
\over\mathrm{lod}(\theta _{\mathrm{ob}}, \theta _0|Y_{\mathrm{ob}})},
\]
where
$S^*_0(Y_{\mathrm{ob}})=\mathrm{E}(S(Y_{\mathrm{co}})|Y_{\mathrm{ob}}, \theta _0)$,
that is, the mean imputation of the missing $S(Y_{\mathrm{co}})$ under the null. Therefore,
$\mathcal{R}I_0$ is
the general version of the second case in~(\ref{rform}), and it
measures the conservativeness of our test when we impute under the
null. Its main disadvantage, as previously mentioned, is that it
can provide very misleading information when the true $\theta $ is far
away from the null. On the other hand, because it is computed at
the null, it is less sensitive, compared to $\mathcal{R}I_1$, to possible
misspecification of the alternative model. We will illustrate this
in Section \ref{subsec:finite}, where we will discuss further the pros
and cons of $\mathcal{R}I_0$.

\begin{figure}[b]

\includegraphics{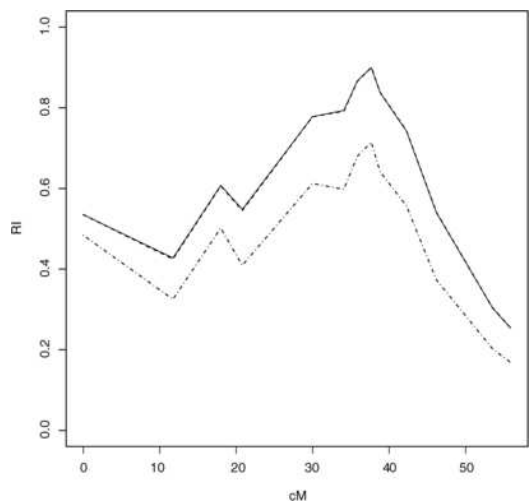}

\caption{The large-sample measures of information
are plotted against the genetic distance. The top two curves
(almost identical) correspond to $\mathcal{R}I_1$ and
$\mathcal{R}I_0$; the
bottom curve (dot-dashed) corresponds to the entropy-based measure
(\protect\cite*{krug96}).\label{fig:niddm}}
\end{figure}

\subsection{Illustration With a Linkage Analysis}

In the context of allele-sharing methods,
the measures we introduced in the previous sections are implemented in
the software ALLEGRO
(\cite*{gudb00}), and are discussed in detail in \citet{nico04}.
In Figure \ref{fig:niddm}, $\mathcal{R}I_1$ and $\mathcal{R}I_0$ are
plotted for
various locations along chromosome 22 (the unit for the X-axis is
CentiMorgans) in a data set consisting of 127 pedigrees used in an
inflammatory bowel disease study (\cite*{cho98}). It can be seen
that, in this case, the two measures are very close across the
entire chromosome. This happens because the sample size is large
and the distribution of the family sharing scores is fairly
symmetric. Also plotted is an inheritance-vector-based information
measure calculated by the software GENEHUNTER (\cite*{krug96}).
This measure takes advantage of the fact that the inheritance
vectors are equally likely under $H_0$ and that, for the fixed
support of the space of the inheritance vectors, the Shannon
entropy (\citeyear{shan49}) is maximal for the uniform
distribution on the support. For the $i$th pedigree in the study and a given
position $t$, it is defined as
\begin{eqnarray*}
&& 1-{E^i(t)\over E_0^i}
\\
&&\quad \equiv1-{-\sum_{\omega_i} P(\omega_i\vert\mathrm{data},
H_0) \log_2 P(\omega_i\vert\hbox{data},H_0)
\over-\sum_{\omega_i} P(\omega_i\vert H_0)\log_2 P(\omega_i\vert H_0)},
\end{eqnarray*}
where $\omega_i$ was defined in Section \ref{sec:gene}.
The definition of the overall
information content of a data set is based on the global entropy,
which, summed over all $n$ pedigrees, satisfies
\[
\mathcal{E}_R=1-{E(t)\over E_0}
\equiv1-{\sum_{i=1}^{n} E^i(t)\over\sum_{i=1}^{n}E_0^i}.
\]
While $\mathcal{E}_R$ has several desired properties (e.g., it is
always between zero and one, and it is one when there is perfect
data on the inheritance vectors), it has some deficiencies that
make it unsuitable for the linkage application. The most
fundamental problem is that it measures the relative information
in the whole inheritance vector space, which could be very
different from what is available for a particular test statistic
that is a function of the inheritance vectors. For example, in the
right diagram of Figure~\ref{fig:sibpair}, we may be nearly
certain, and hence suffer very little missing information, that
the IBS sharing is actually IBD if we have the knowledge that the
allele ``A2'' has very low population frequency, even though the
parental alleles are unknown and therefore $\mathcal{E}_R$ is low
(see \cite*{nico04}, for more details). It is also possible that
$\mathcal{E}_R$ is higher than the measures described in this
paper (e.g., \cite*{thala05}), for example in situations where
there is a lot of data on unaffected individuals in a family, but
little or no data on affected individuals. In these cases,
$\mathcal{E}_R$ will capture available information that is not
directly of interest when we are performing affecteds-only
analyses.

\begin{figure*}

\includegraphics{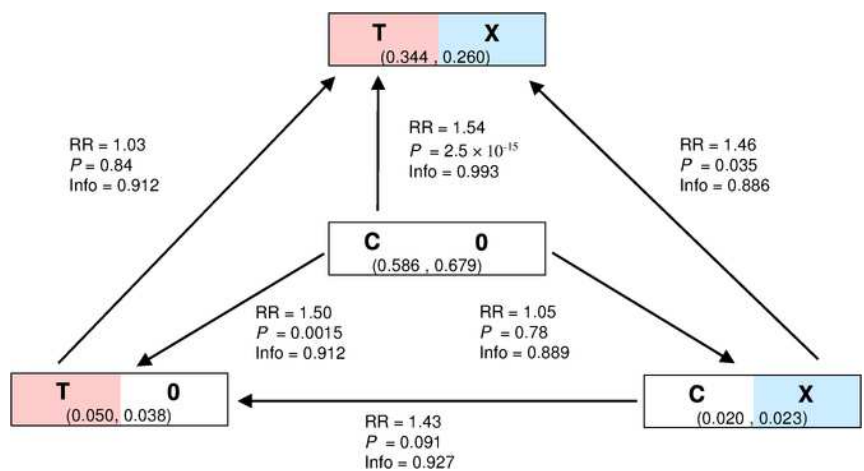}

\caption{For each haplotype, estimated frequencies in patients and controls are
displayed. RR
is estimated risk of the haplotype the arrow is pointing to ($h_1$)
relative to the
haplotype the arrow is pointing from ($h_2$), and is calculated as
$[n(h_1)/m(h_1)]/[n(h_2)/m(h_2)]$ where n and m are estimated haplotype
counts in patients
and controls respectively. P values are calculated based on a
likelihood ratio test
that properly takes missing information into account. Information shown
is $\mathcal{R}I_1$. Very similar numbers are obtained
for $\mathcal{R}I_0$.\label{fig:stroke}}
\end{figure*}

The serious overestimation or underestimation of relative
information can have a great impact on the design of follow-up
studies. One can decide on increasing the marker density if the
relative information is low, as opposed to increasing the sample
size. Both strategies are expensive, and therefore deciding what
is the most efficient design is of great importance in practice.
For example, at the global mode in Figure~\ref{fig:niddm}, our
measures indicate that we have about $90\%$ relative information,
implying that potentially we can increase the lod score by only
about $11\%$ ($1/0.9=1.11$) if we add markers to make the IBD
process approximately known
(assuming the value of $\theta _{\mathrm{ob}}$ remains
approximately the same with the additional data). On the other
hand, the entropy-based measure from GENEHUNTER indicates that we
have about 70\% information, suggesting that a more substantial
gain (over $40\%$) is possible by increasing the density of the
markers. Therefore these two approaches of measuring information
are likely to lead to different strategies in allocating the
resources, but evidently, in this example, it is unlikely the
test results will change significantly by adding more markers near
the location at the global mode.

\subsection{Illustration With a Haplotype-Based Study}\label{s4.5}

In \citet{grant06}, the gene \emph{TCF7L2} was found to be
associated with type-2 diabetes. In particular, allele~T of
\textit{rs7903146} (SNP402) and allele X of a microsatellite marker
DG10S478 are both associated with elevated risk of type-2 diabetes
($p$-value${}<10^{-10}$). Allele T and allele X are substantially
correlated ($r \approx0.85$) and their effects could not be
clearly distinguished from each other in the original study.
However, with additional data (\cite*{helgason07}), it became
clear that allele T is more strongly associated with diabetes than
allele X. SNP402 has alleles T and C, and DG10S478 has alleles X
and 0. Jointly there are four haplotypes: TX, CX, T0 and C0.
Figure \ref{fig:stroke} presents pairwise comparisons of these
four haplotypes. Data are from 1021 patients ($n = 2042$
chromosomes) and 4273 controls ($m = 8546$ chromosomes).
Consistent with the single marker associations, haplotype TX is
found to have elevated risk relative to C0. To distinguish between
the effects of alleles T and X, haplotype T0 is found to confer
risk that is similar to that of TX and has significantly higher risk
than C0. By contrast, haplotype CX is found to have risk similar
to that of C0 and significantly lower risk than TX. In other
words, given SNP402, DG10S478 does not appear to provide extra
information about diabetes, which keeps SNP402 as a strong
candidate for being the functional variant.

The yield of the genotypes is not perfect. Each subject has
genotypes for at least one of the two markers, but about 3.5\% of
the genotypes are missing. This together with uncertainty in phase
leads to the incomplete information summarized in Figure
\ref{fig:stroke}. Interestingly, while the same data are used for
the six pairwise comparisons, the fraction of missing information
can be quite different. Most striking is that the relative
information for the test of TX versus C0 is very close to 100\%,
while the other tests all have more substantial missing
information. We explore some of the reasons below.

Notice that T is highly correlated with X and C highly correlated with
0. As a
consequence, TX and C0 are much more common than T0 and CX. Consider a
subject whose
genotype for D10GS478 is missing. Here we can think of his two alleles
for SNP402
one at a time. Given an observed allele T, it is clear that the
haplotype is not C0
and quite likely to be TX. Hence, even though incomplete, there is
still substantial
information provided for the test of TX versus C0. By contrast, we know
that this
chromosome is useful for the test of TX against T0, but with the allele
of DG10S478
missing, that information is completely lost. Even more interesting is
that, if the
observed allele is C instead, then this haplotype is completely
uninformative for
the test of TX versus T0, that is, there is actually no information here
whether or not
we know the corresponding DG10S478 allele. In effect, the genotype of
SNP402 is an
ancillary statistic for the test of TX against T0 (or CX against C0).
It tells us
how much information we can get from this individual assuming that we
have no
missing data, but by itself does not provide any information for the
test. Moreover,
if the test of TX versus T0 is of key interest, then effort to fill up missing
genotypes for DG10S478 should be focused on those individuals that are T/T
homozygous for SNP402.

When genotypes of both markers are observed, uncertainty in phase
only exists for those individuals that are doubly heterozygous,
that is, having genotypes C/T and 0/X. Such an individual either has
haplotypes C0/TX (scenario I) or CX/T0 (scenario II). Scenario II
provides no information for the test of TX versus C0. Scenario I
does contribute something to the test, but by providing a count of
1 to both TX and C0, its impact on the test of TX versus C0 is
rather limited. By contrast, for the test of TX versus T0,
scenario I adds a count of 1 to TX while scenario II adds a count
to T0. Hence, uncertainty in phase has a much bigger impact on the
test of TX versus T0 than the test of TX versus C0. This example,
therefore, illustrates clearly the importance of measuring
\emph{test-specific} relative information.

\section{ Small-sample Exploratory Measures}\label{sec:small}

\subsection{A Bayesian Framework}\label{subsec:bayes}

The measures defined in previous sections do not necessarily work
with small samples (e.g., data for one family) because they rely
on the ability of the MLE to summarize the whole likelihood
function. The Bayesian approach becomes a valuable tool in such
settings even if we do not necessarily have a reliable prior; we
can first construct a coherent measure and then investigate the
choice of prior. Since a likelihood quantifies the information in
the data through its ability of distinguishing different values of
the parameter, it is natural to consider measuring the relative
information by comparing how the observed-data likelihood deviates
from ``flatness'' relative to the same deviation in the
complete-data likelihood. The Bayesian method is ideal here
because we need to assess the change in this deviation due to the
joint variability in the missing data and in the parameter. A
reasonable measure of this deviation, conditioning on $Y_{\mathrm{ob}}$, is
the posterior variance of the likelihood ratio (LR). This measure
is appealing because it is naturally scaled via the equality
%
\begin{equation}\label{ratio}
\quad
\mathrm{LR}(\theta _0, \theta |Y_{\mathrm{ob}}) =\mathrm{E}
[\mathrm{LR}(\theta _0, \theta |Y_{\mathrm{co}})
|Y_{\mathrm{ob}}, \theta  ],
\end{equation}
which guarantees that
%
\begin{equation}\label{ripi}
0\le\mathcal{B}I^\pi _1 \equiv\frac{\operatorname{Var}
[\mathrm{LR}(\theta _0, \theta  |Y_{\mathrm{ob}})|Y_{\mathrm{ob}} ]}
{\operatorname{Var} [\mathrm{LR}(\theta _0, \theta|
Y_{\mathrm{co}})|Y_{\mathrm{ob}} ]} \le1,
\end{equation}
where $\pi$ indexes the underlying prior on $\theta$ used
by~(\ref{ripi}), and $\mathcal{B}I$ stands for ``Bayes Information.'' We
assume here that the complete-data likelihood surface is not flat,
as otherwise the model is of little interest. The denominator in
(\ref{ripi}) is therefore positive. We also need to assume that
the posterior variances of the two likelihood ratios are finite.
This second assumption can be violated in practice, but a second
measure we will propose below essentially circumvents this
problem.

In the presence of nuisance parameters (under the null), there is
also a subtle issue regarding the nuisance part of $\theta _0$, in
the definition of $\mathcal{B}I^\pi _1$. For a full Bayesian
calculation, one
should leave it unspecified and average it over in the posterior
calculation, just as with the $\theta $ in
$\mathrm{LR}(\theta _0,\theta )$. On the
other hand, to be consistent with the large-sample measures as
defined in Section~\ref{sec:large}, we can fix the nuisance
parameter part in $\theta _0$ by its observed-data MLE under the
null. Identity~(\ref{ratio}) still holds with such a ``fix,''
because the calculation there conditions on the observed data.
This ``fix'' may seem to be rather ad hoc from a pure Bayesian
point of view. However, it can be viewed as an attempt in
capturing the dependence (if any) between the parameter of
interest and the nuisance parameter under the null, a dependence
that is ignored by a single prior on the nuisance parameter
regardless of the null. This subtle issue is related to the
difference between ``estimation prior'' and ``hypothesis testing
prior,'' an issue we will explore in subsequent work. Here we just
note that all the Bayesian measures defined in this section can be
constructed with either approach for handling the nuisance
parameter under the null, although those under shrinking prior
toward the null (see Section~\ref{subsec:shrink}) are most easily
obtained when the nuisance parameter under the null is fixed at
its MLE (or some other known values).

With either approach,
\begin{eqnarray*}
&& \mathcal{B}I^\pi _1 =1 \quad\mbox{if and only if }
\\
&&\quad
\mathrm{E} \{\operatorname{Var} [\mathrm{LR}(\theta _0,
\theta  |Y_{\mathrm{co}})|Y_{\mathrm{ob}}, \theta
] |Y_{\mathrm{ob}} \} =0,
\end{eqnarray*}
which occurs if and only if for almost all the $\theta $ in the
support of the posterior, the complete-data likelihood
$\mathrm{LR}(\theta _0, \theta  |Y_{\mathrm{co}})$ is (almost
surely) a constant as a
function of the missing data, and thus the missing data would
offer no additional help in distinguishing $\theta $ from~$\theta_0$. On
the other hand, $\mathcal{B}I^\pi _1 =0$ if and only if the observed-data
likelihood ratio is a constant, and thus there is no information
in the observed data for testing $H_0$ using $\mathrm{LR}(\theta _0,
\theta  |Y_{\mathrm{ob}})$. Other characteristics of this measure
depend on the
choice of the prior $\pi$, and they will be discussed in the
following sections.

One potential drawback of $\mathcal{B}I^\pi _1$ is that it can be greatly
affected by the large variability in the likelihood ratios, as
functions of the parameters, for example, when very unlikely
parameter values were given nontrivial prior mass. This problem
can be circumvented to a large extent by using the posterior
variance of the \textit{log-likelihood ratio},
\[
\operatorname{Var} [\mathrm{lod}(\theta , \theta _0
|Y_{\mathrm{ob}})|Y_{\mathrm{ob}} ].
\]
The use of the log scale also makes it much more likely, compared
to the ratio scale, that the resulting posterior variances are
finite. Evidently, just as with the posterior variance of the
likelihood ratio, this is equal to zero if and only if the
observed-data likelihood ratio is a constant (on the support of
the posterior). Similarly,
\[
\operatorname{Var} \biggl[  \log\frac{P(Y_{\mathrm{co}}|
Y_{\mathrm{ob}},\theta )}{P(Y_{\mathrm{co}}|
Y_{\mathrm{ob}},\theta _0)}\bigg | Y_{\mathrm{ob}} \biggr]
\]
is equal to zero if and only if there is no additional information
in the missing data for testing $H_0$. These suggest that we can
also measure the relative information by
%
\begin{eqnarray}\label{bayes-ri}
\qquad
\mathcal{B}I^\pi _2 &=& \operatorname{Var}
[\mathrm{lod}(\theta , \theta _0 |Y_{\mathrm{ob}})|Y_{\mathrm{ob}} ]
\nonumber \\
&&{}\cdot \biggl(\operatorname{Var} [\mathrm{lod}(\theta , \theta _0|
Y_{\mathrm{ob}})|Y_{\mathrm{ob}} ]
\\
&&\hspace*{13pt}{}
+ \operatorname{Var} \biggl[\log{\frac{P(Y_{\mathrm{co}}|
Y_{\mathrm{ob}},\theta )}{P(Y_{\mathrm{co}}|Y_{\mathrm{ob}},
\theta_0)}}\Big|Y_{\mathrm{ob}} \biggr]\biggr)^{-1},\hspace*{-4pt}
\nonumber
\end{eqnarray}
where, as for $\mathcal{B}I^\pi _1$, $\pi$ indexes the underlying
prior on~$\theta$.

Although the use of lod is more natural in view of the
large-sample measures given in Section~\ref{sec:large}, it does
not admit the nice ``coherence'' identity for the likelihood ratio
as given in (\ref{ratio}). Indeed, we had to remove ad hoc a
cross term in the denominator of (\ref{bayes-ri}) in order to keep
the resulting ratio always inside the unit interval.
Furthermore, as we show in Section~\ref{sec:theor}, the use of the
ratio scale, instead of log ratio, leads to a number of
interesting identities between likelihood ratios and Bayes
factors, and it is more connected with some finite-sample measure
of information in the literature. Whereas such trade-offs need to be
explored, our general results in the next section imply that in the
neighborhood of $\theta _0$, the differences between these two
measures should be small.

\subsection{Limits Under a Shrinking Prior Toward Null}\label{subsec:shrink}

Given their definitions, the immediate question is how to choose
$\pi$ and how to compute $\mathcal{B}I^\pi _1$ and $\mathcal{B}I^\pi _2$ efficiently since,
in general, their calculations require integrations that cannot be
performed analytically. When the truth is believed to be in a
neighborhood of the null value $\theta _0$, a $\theta _0$-neighbor
approximation to $\mathcal{B}I^\pi _1$ and $\mathcal{B}I^\pi _2$
can be obtained by
choosing $\pi$ to be $U(\theta _0-\delta, \theta _0 + \delta)$ with
$\delta> 0$ small. It is proved in Appendix~\ref{sA.1} that the two
Bayesian measures have the same limit as $\delta\rightarrow0$,
denoted by $\mathcal{B}I_0$,
%
\begin{eqnarray}\label{eq:bisg}
\qquad
\mathcal{B}I_0 &=&\frac{ S^2(\theta _0|Y_{\mathrm{ob}}) }
{S^2(\theta _0|Y_{\mathrm{ob}})
+ \operatorname{Var}(S(\theta _0|Y_{\mathrm{co}})|
Y_{\mathrm{ob}},\theta _0) }
\nonumber\\[-8pt]
\\[-8pt]
&=& \frac{ S^2(\theta _0|Y_{\mathrm{ob}}) }
{S^2(\theta _0|Y_{\mathrm{ob}}) +I_{\mathrm{mi}}
(\theta_0|Y_{\mathrm{ob}}) },
\nonumber
\end{eqnarray}
where $S(\theta |Y_{\mathrm{ob}})$ and $S(\theta |Y_{\mathrm{co}})$
are respectively the
observed-data and complete-data score function, and $I_{\mathrm{mi}}(\theta |Y_{\mathrm{ob}})$
is the expected (missing) Fisher information from
$f(Y_{\mathrm{co}}|Y_{\mathrm{ob}},\theta )$. Note that although
this result obviously assumes
$\theta $ is univariate, it can also be applied when only the
parameter of interest is univariate, if we fix the nuisance
parameter part in $\theta _0$ at its observed-data MLE under the
null.

For the exponential tilting linkage model, one can verify that
%
\begin{eqnarray}\label{eq:bisp}
\mathcal{B}I_0 &=& \frac{W^2 } {W^2 + \operatorname{Var}(Z|\mathrm{data}, H_0)}
\nonumber\\[-8pt]
\\[-8pt]
&=& 1-\frac{\operatorname{Var}(Z|\hbox{data}, H_0) }
{W^2 + \operatorname{Var}(Z|\hbox{data}, H_0)},
\nonumber
\end{eqnarray}
where $W=\mathrm{E}(Z|\hbox{data}, H_0)$, and $Z$ is given in (\ref{NPL}).
Therefore its computation is straightforward because it only
depends on the test statistic and the null hypothesis. Note also
that the expectation of the denominator in (\ref{eq:bisp}) under
the null is simply $\operatorname{Var}(Z|H_0)=1$. Therefore, if we
replace the
denominator in (\ref{eq:bisp}) by its expected value under the
null, we obtain an even simpler approximation
$\mathcal{B}I_0\approx1 -\operatorname{Var}(Z|\hbox{data}, H_0)$.

\begin{figure}

\includegraphics{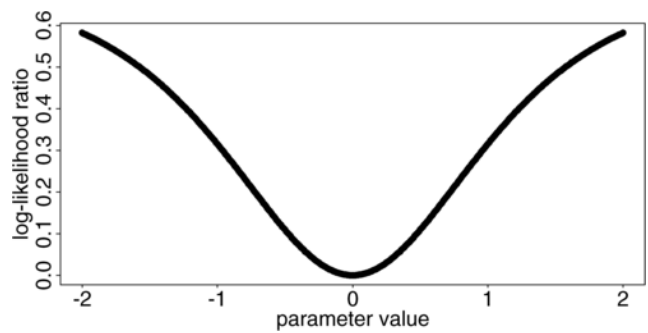}

\caption{Log-likelihood ratio for a sib-pair
where the parents and sibs are IBS for a heterozygous genotype.\label{fig:dh-loglik}}
\end{figure}

However, $\mathcal{B}I_0$ measures only
the relative information in the neighborhood of $\theta _0$. For
example, suppose the data consist of one affected sib-pair like in
Figure~\ref{fig:sibpair} such that both parents and the sibs are
heterozygous with the same pair of alleles at a specific locus
(i.e., all individuals have the alleles ``A1'' and ``A2''). In this
case, the observed-data likelihood from the exponential tilting
model is very informative away from $\theta _0$ (see Figure
\ref{fig:dh-loglik}), but $\mathcal{B}I_0=0$ because the null value
$\theta _0=0$ turns out to be the \textit{minimizer} of the
observed-data likelihood.

In general, whenever $\theta _0$ is a stationary point of
$\ell(\theta |Y_{\mathrm{ob}})$, $\mathcal{B}I_0=0$, even if there
is almost perfect
information. For example, if the data consist of $2n+1$ sib-pairs
such that there is complete information on $2n$ sib-pairs, $n$
sharing 0 alleles IBD and $n$ sharing 2 alleles IBD, and one
sib-pair has no information, then $W=0$ and thus $\mathcal{B}I_0=0$. This
is clearly a misleading measure. In the next section we propose a
remedy for this problem.

\subsection{Combining Individual Information Measures}\label{subsec:comb}

The measures defined in Section~\ref{subsec:bayes} are inherently
small-sample quantities, for the variance terms used in these
measures do not naturally admit additivity even for i.i.d. data
structures. Whether one can find a satisfying small-sample measure
that would automatically admit such additivity is a topic of both
theoretical and practical interest, but for our current purposes
we can impose such additivity by defining global measures via
appropriate combining rules, such as (\ref{eq:comb}). We adopt
such rules mainly to maintain the continuity of moving from
small-sample to large-sample measures as proposed in
Section~\ref{sec:large}. Whether these are the most sensible rules
is a topic that requires further research.

Specifically, suppose our data consist of $n$ independent ``small
units'' (e.g., individual families), $Y^{(i)}_{\mathrm{ob}}$. We apply
(\ref{ripi}) to each unit and then combine them via the harmonic
rule (\ref{eq:comb}) but with weights proportional to $V_i\equiv
\operatorname{Var} [\mathrm{LR}(\theta _0, \theta
|Y_{\mathrm{ob}}^{(i)})|Y_{\mathrm{ob}}^{(i)} ]$. In other
words, we define the measure for the aggregated data by first
summing up both the numerators and denominators of individual
$\mathcal{B}I^\pi _{1,i}$ and then taking the ratio. That is,
%
\begin{eqnarray}\label{ripin}
\mathcal{B}I^\pi _1
&=& \frac{\sum_{i=1}^n \operatorname{Var}[\mathrm{LR}(\theta _0,
\theta  |Y_{\mathrm{ob}}^{(i)})|Y_{\mathrm{ob}}^{(i)} ]}
{\sum_{i=1}^n\operatorname{Var} [\mathrm{LR}(\theta _0,
\theta|Y_{\mathrm{co}}^{(i)})|Y_{\mathrm{ob}}^{(i)} ]}
\nonumber\\[-8pt]
\\[-8pt]
&=& \biggl\{\frac{\sum_{i=1}^n V_i[\mathcal{B}I^\pi _{1,i}]^{-1}}
{\sum_{i=1}^nV_i} \biggr\}^{-1}.
\nonumber
\end{eqnarray}

Similarly, we can define the combined version for $\mathcal{B}I^\pi
_2$ from
individual $\mathcal{B}I^\pi _{2,i}$, and we can also use the arithmetic
combining rule (\ref{eq:combl}). In addition, its limit under the
shrinking prior, in analogy to (\ref{eq:bisg}), can be expressed
as
%
\begin{eqnarray}\label{eq:combss}
\qquad
\mathcal{B}I_s & = & \frac{\sum_{i=1}^n S^2
(\theta_0|Y^{(i)}_{\mathrm{ob}})}
{\sum_{i=1}^n S^2(\theta _0|Y^{(i)}_{\mathrm{ob}})
+ \sum_{i=1}^n I_{\mathrm{mi}}(\theta _0|Y_{\mathrm{ob}}^{(i)})}
\nonumber\\[-8pt]
\\[-8pt]
& = & \frac{\sum_{i=1}^n S^2(\theta _0|Y^{(i)}_{\mathrm{ob}})}
{\sum_{i=1}^n S^2(\theta _0|Y^{(i)}_{\mathrm{ob}})
+ I_{\mathrm{mi}}(\theta_0|Y_{\mathrm{ob}})},
\nonumber
\end{eqnarray}
where $I_{\mathrm{mi}}(\theta |Y_{\mathrm{ob}})$ is the expected
Fisher information matrix
from $f(Y_{\mathrm{co}}|Y_{\mathrm{ob}}, \theta )$, with
$Y_{\mathrm{ob}}=\{Y_{\mathrm{ob}}^{(1)},\ldots,\break  Y_{\mathrm{ob}}^{(n)}\}$.
We have changed the notation
from $\mathcal{B}I_0$ to $\mathcal{B}I_s$ to signify the fact that
the latter
measure is defined by \textit{summing} up the numerators and
denominators of the individual $\mathcal{B}I_0$'s \textit{separately} before
forming the combined ratio. The second equation in
(\ref{eq:combss}) holds because of the additivity of Fisher
information for independent data structures. For the exponential
tilting linkage model, this averaging for a shrinking prior leads
to
\begin{eqnarray*}
\mathcal{B}I_s &=& \frac{\sum_{i=1}^n W_i^2 }
{\sum_{i=1}^n W_i^2+\sum_{i=1}^n \operatorname{Var}(Z_i|\mathrm{data},H_0)}
\\
&=& \frac{\sum_{i=1}^n W_i^2/n}
{\sum_{i=1}^n W_i^2/n +\operatorname{Var}(Z|\mathrm{data},H_0)},
\end{eqnarray*}
where $W_i=\mathrm{E}(Z_i|\hbox{data}, H_0)$
and $Z=\sum_{i=1}^n Z_i/\sqrt{n}$. This is equal to zero only if
all the $W_i$'s are equal to zero, as opposed to using a global
posterior, that is, by applying (\ref{eq:bisg}) directly to the
whole data set, where $\sum W_i=0$ is sufficient to cause
$\mathcal{B}I_0=0$. This difference is an important advantage for
$\mathcal{B}I_s$,
as we will demonstrate in Section~\ref{subsec:finite}.

\begin{figure}

\includegraphics{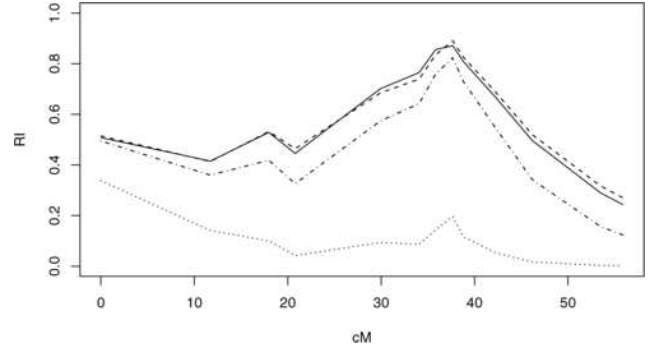}

\caption{The Bayesian measures are calculated for a
data set containing 21 families. The solid line is $\mathcal{B}I_s$; the
dashed line corresponds to $\mathcal{B}I^\pi _2$ calculated using a uniform
prior on $(-1,1)$; the dot-dashed line corresponds to $\mathcal{B}I^\pi _2$
calculated using a uniform prior on
$(\min(\theta _{\mathrm{ob}},\theta _0)-0.1,\max(\theta _{\mathrm{ob}},\theta _0)+0.1)$; the dotted line
corresponds to $\mathcal{B}I^\pi _2$ calculated using a uniform prior on
$(\theta _{\mathrm{ob}}-0.1,\theta _{\mathrm{ob}}+0.1)$.\label{fig:buc}}
\end{figure}

\subsection{An Empirical Comparison}\label{sec:exam}

To illustrate the proposed Bayesian measures of information, we
calculated them for various priors $\pi$ in a data set containing
21 ulcerative colitis (UC) families (\cite*{cho98}). The choices
of priors here were made for investigating the sensitivity to
prior specification, so they may not reflect our real knowledge
about the problem (e.g., we generally expect $\theta$ to be
nonnegative in such problems). In Figure \ref{fig:buc} the
measure of information $\mathcal{B}I_s$ is plotted in comparison with
$\mathcal{B}I^\pi _2$, which is calculated as described in the previous
section for three different priors. Similar results are obtained
using $\mathcal{B}I^\pi _1$. In this example $\mathcal{R}I_1$ and
$\mathcal{B}I_s$ are almost
identical; $\mathcal{R}I_1$ is therefore not shown. Note that the
value of
the parameter under the null hypothesis of no linkage is equal to
zero, and, for this data set, the maximum likelihood estimates for
the linkage parameter across the chromosome vary between $-0.74$
and~0.07.\looseness=-1

We note that the $\mathcal{B}I^\pi _2$ measure calculated using a
Uniform$(-1,1)$ prior is very close to $\mathcal{B}I_s$, which demonstrates
the possibility of having very different priors that result in
very similar measures. The Bayesian measure calculated with a
prior having a narrower support, that is, uniform on the interval
$(\min(\theta _{\mathrm{ob}},\theta _0)-0.1,\max(\theta _{\mathrm{ob}},\theta _0)+0.1)$,
follows the same patterns but is uniformly smaller. Using a prior centered
around the maximum likelihood estimate, uniform on the interval
$(\theta _{\mathrm{ob}}-0.1,\theta _{\mathrm{ob}}+0.1)$, turns out
to be very misleading because
it gives values that are considerably too small (i.e., in
comparison with the large-sample estimates given in
Figure~\ref{fig:niddm}). We emphasize that symmetric uniform
priors were used in Figure~\ref{fig:buc} simply to demonstrate
potential substantial sensitivity to prior specification, as one
often expects less erratic behavior from such symmetric and smooth
prior specifications. The issue of sensitivity to the choice of
prior is further discussed in Section~\ref{sec:future}.

\section{Theoretical Connections, Comparisons and
Curiosities}\label{sec:theor}

\subsection{The Asymptotic Equivalence to the Estimation~Measure}\label{subsec:asym}

As we discussed previously, a central difficulty in measuring the
relative amount of information is that its value will generally
depend on the true value of the unknown parameter. One way to
explore this dependence is to replace $\theta _{\mathrm{ob}}$ in the
definition of
$\mathcal{R}I_0$ or $\mathcal{R}I_1$ by $\theta $ in a suitably
defined neighborhood,
and to plot it against $\theta $ in such a range to check its
variability. The use of this type of \textit{relative information
function} was proposed in \citet{mengvand96} for the purpose of
measuring the rate of convergence of EM-type algorithms, where
the function
%
\begin{equation}\label{fracfu}
\mathcal{R}I(\theta ) = {\ell_{\mathrm{ob}}(\theta _{\mathrm{ob}})
- \ell_{\mathrm{ob}}(\theta )\over Q(\theta _{\mathrm{ob}}|
\theta_{\mathrm{ob}}) - Q(\theta |\theta _{\mathrm{ob}})}
\end{equation}
was termed \textit{relative augmentation function}. Note that
$\mathcal{R}I_1$ is simply the value of this function at $\theta
=\theta _0$. For
simplicity of presentation, we will assume in the following and
Section~\ref{subsec:fixed-size} that $\theta $ is univariate, though
all the results are generalizable to multivariate $\theta $ by
employing appropriate matrix notation and operations. We also
assume all the regularity conditions as in
Dempster, Laird and Rubin (\citeyear{demplairrubi77}) 
to guarantee the validity of taking
differentiation under integration and for Taylor expansions.

It was shown in \citet{mengvand96} that as
$\theta\rightarrow\theta _{\mathrm{ob}}$, $\mathcal{R}I(\theta )$
approaches the so-called
\textit{fraction of observed information} for the purpose of estimation:
%
\begin{equation}\label{eq:dlr}
\mathcal{R}I_{E} = {I_{\mathrm{ob}}\over I_{\mathrm{co}}}
\equiv {I_{\mathrm{ob}}\over I_{\mathrm{ob}}+ I_{\mathrm{mi}}},
\end{equation}
where the observed, complete and missing Fisher information are
defined, as in Dempster, Laird and Rubin (\citeyear{demplairrubi77}), 
%
\begin{eqnarray}
\qquad \quad
I_{\mathrm{ob}}&\equiv &I_{\mathrm{ob}}(\theta _{\mathrm{ob}})
= -\frac{\partial^2 \log f(Y_{\mathrm{ob}}|\theta)}
{\partial\theta ^2} \bigg\vert_{\theta =\theta _{\mathrm{ob}}},
\label{eq:iobs}\\
I_{\mathrm{mi}} &\equiv & I_{\mathrm{mi}}(\theta _{\mathrm{ob}})
\nonumber\\[-8pt]\label{eq:imis}
\\[-8pt]
&=& \mathrm{E} \biggl[-\frac{\partial^2 \log f(Y_{\mathrm{co}}|Y_{\mathrm{ob}};
\theta )}{\partial\theta ^2} \bigg|Y_{\mathrm{ob}}; \theta\biggr]
\Big\vert_{\theta =\theta_{\mathrm{ob}}}
\nonumber
\end{eqnarray}
and
%
\begin{eqnarray}\label{eq:icom}
I_{\mathrm{co}}&\equiv & I_{\mathrm{co}}(\theta _{\mathrm{ob}})
\nonumber \\
&=& \mathrm{E} \biggl[-\frac{\partial^2 \log
f(Y_{\mathrm{co}}|\theta )}{\partial\theta ^2} \Big|Y_{\mathrm{ob}};
\theta  \biggr] \bigg\vert_{\theta =\theta _{\mathrm{ob}}}
\\
&=& I_{\mathrm{ob}}+I_{\mathrm{mi}},
\nonumber
\end{eqnarray}
where the last identity is known as the ``missing-data principle,''
and is a directed consequence of (\ref{eq:emex}). The $\mathcal{R}I_E$
measure plays a key role in determining the rate of convergence of
the EM algorithm and its various extensions
(e.g., \cite*{demplairrubi77};
\cite*{mengrubi91}, \citeyear{mengrubi93};
\cite*{meng94ecm};
\cite*{mengvand97}).

The above limiting result suggests that, when $\delta=\theta _0 -
\theta _{\mathrm{ob}}$ is small, we can study the behavior of
$\mathcal{R}I_1$ via its
connection to $\mathcal{R}I_{E}$, as we demonstrate in the next section.
However, among all the measures we proposed, the measure $\mathcal{B}I_s$
of (\ref{eq:combss}) most closely resembles $\mathcal{R}I_{E}$ of
(\ref{eq:dlr}). The main differences are the use of $\sum_{i=1}^n
S^2(\theta _0|Y_{\mathrm{ob}}^{(i)})$ in place of $I_{\mathrm{ob}}$,
and the fact that the
Fisher information terms in $\mathcal{R}I_{E}$ are evaluated at
$\theta =\theta _{\mathrm{ob}}$, whereas for $\mathcal{B}I_s$ they
are evaluated at $\theta =\theta _0$. It is well known
that, under regularity conditions, $\sum_{i=1}^n
S^2(\theta _0|Y_{\mathrm{ob}}^{(i)})/n$ will converge to the expected Fisher
information under the null. Consequently, under the null, $\mathcal{B}I_s$
and $\mathcal{R}I_E$ are asymptotically equivalent. This equivalence may
suggest to directly define $\mathcal{B}I_s$ in terms of the ``observed
Fisher information at $\theta _0$.'' However, although
$I_{\mathrm{ob}}\equiv I_{\mathrm{ob}}(\theta _{\mathrm{ob}})$
is guaranteed to be nonnegative (definite) when
$\theta _{\mathrm{ob}}$ is in the interior of the parameter space
$\Theta$, this
is not necessarily true for $I_{\mathrm{ob}}(\theta _0)$.
Therefore, for small-sample problems for which
the use of $I_{\mathrm{ob}}$ is inadequate (e.g., when the MLE
$\theta _{\mathrm{ob}}$ is on
the boundary of $\Theta$), the direct substitution of
$I_{\mathrm{ob}}$ by $I_{\mathrm{ob}}(\theta _0)$ will not lead, in general, to a
nonnegative measure.
The $\mathcal{B}I_s$ measure \mbox{circumvents} this problem by using the sum of
individual squared scores instead of $I_{\mathrm{ob}}(\theta _0)$, which
guarantees that the resulting measure is inside the unit
interval, and that it is consistent with $\mathcal{R}I_E$ for large
samples. Therefore $\mathcal{B}I_s$ can be viewed as a small-sample
extension of $\mathcal{R}I_E$ in the neighborhood of the null.

\subsection{Finite-Sample Equivalence in the Neighborhood of
the Null}\label{subsec:fixed-size}

For both $\mathcal{R}I_1$ and $\mathcal{R}I_0$, their equivalence to
$\mathcal{R}I_E$ in
the neighborhood of $\theta _0$ can be established for finite-sample
sizes. (Therefore, $\mathcal{R}I_E$ can also be defined as the value of
either $\mathcal{R}I_1$ or $\mathcal{R}I_0$ when
$\theta _{\mathrm{ob}}=\theta _0$.) Specifically,
denote $\ell^{(k)}_{\mathrm{ob}}$ the $k$th derivative of
$\ell_{\mathrm{ob}}(\theta )$ at $\theta =\theta _{\mathrm{ob}}$, and
%
\begin{equation}\label{eq:dq}
Q^{(i,j)}_{\mathrm{ob}}\ = \frac{\partial^{i+j}Q(\theta _1|\theta _2)}
{\partial\theta ^i_1\,\partial\theta ^j_2}
\bigg\vert_{\theta_1=\theta _2=\theta _{\mathrm{ob}}}.
\end{equation}
It is proved in Appendix \ref{sA.2} that
%
\begin{equation}\label{eq:rione}
\qquad
\mathcal{R}I_1= \mathcal{R}I_{E} + \frac{Q^{(3,0)}_{\mathrm{ob}}
\mathcal{R}I_E - \ell^{(3)}_{\mathrm{ob}}}{3I_{\mathrm{co}}}
\delta+ O(\delta^2).
\end{equation}
In deriving this result, we
have utilized the following well-known identities in the
literature of the EM algorithm (e.g., Dempster, Laird and Rubin, \citeyear{demplairrubi77}; 
\cite*{mengrubi91}):
%
\begin{equation}\label{eq:emid}
Q^{(1,0)}_{\mathrm{ob}}=\ell^{(1)}_{\mathrm{ob}}= 0;\quad
Q^{(2,0)}_{\mathrm{ob}}=-I_{\mathrm{co}}.
\end{equation}

Under the assumption that $Q(\theta |\theta _0)$ has a unique maximizer
as a function of $\theta $, an assumption that is easily satisfied in
most of the applications when EM is useful, we also prove in
Appendix~\ref{sA.2} that
%
\begin{eqnarray}\label{eq:rizer}
\quad
\mathcal{R}I_0 &=& \mathcal{R}I_E
\nonumber \\
&&{}  + \bigl(3I_{\mathrm{ob}}\bigl(Q^{(3,0)}_{\mathrm{ob}}
+ Q^{(2,1)}_{\mathrm{ob}}\bigr)
\nonumber\\[-8pt]
\\[-8pt]
&&\hspace*{15pt}{}
- 2\ell^{(3)}_{\mathrm{ob}}
- Q^{(3,0)}_{\mathrm{ob}}\mathcal{R}I_E^2\bigr)
(3I_{\mathrm{co}})^{-1} \delta
\nonumber \\
&&{} + O(\delta^2).
\nonumber
\end{eqnarray}
These expansions are useful for comparing the first-order (in
$\delta$) behavior of $\mathcal{R}I_1$ and $\mathcal{R}I_0$. For
example, we
suspect that, for many applications, $\mathcal{R}I_0$ is a conservative
estimate of the actual relative information, where $\mathcal{R}I_1$ is a
more accurate measure. One way to validate this or to identify
situations where this conjecture is true is to compare the two
coefficients of $\delta$ and to determine the appropriate
conditions for $\mathcal{R}I_0<\mathcal{R}I_1$ to the first order in the
neighborhood of $\theta _0$ (away from the neighborhood the
comparison is not very meaningful because $\mathcal{R}I_0$ can be
seriously biased). Due to the complex nature of these two
coefficients, we only present in the next section a simple example
to illustrate the conservatism of $\mathcal{R}I_1$, and leave the general
theoretical investigation to subsequent work.

We also remark here that when the true $\theta $ is believed to be close
to $\theta _0$, a measure like $\mathcal{R}I_0$ can be used to
construct reasonable bounds. For example, we can expect
$\min\{\mathcal{R}I_0, \mathcal{R}I_1\}$ to be a reasonable lower
bound and
$\max\{\mathcal{R}I_0, \mathcal{R}I_1\}$ an upper bound for relative
information,
or we can use $\mathcal{R}I_{0.5}\equiv\break \sqrt{\mathcal{R}I_0\mathcal{R}I_1}$ as a
compromise. In future work, we intend to investigate the reliability and
applicability of such bounds and compromise. Here we simply note a
computational advantage of $\mathcal{R}I_{0.5}$ that follows from
%
\begin{equation}\label{eq:rico}
\quad
\mathcal{R}I_{0.5} = \biggl[\frac{\max_{\theta }[Q(\theta |\theta _0)
- Q(\theta _0|\theta _0)]}{Q(\theta _{\mathrm{ob}}|
\theta _{\mathrm{ob}})-Q(\theta _0|\theta _{\mathrm{ob}})} \biggr]^{1/2},
\end{equation}
which avoids entirely the calculation of the observed-data
log-likelihood function $\ell_{\mathrm{ob}}(\theta )$, which is
often harder to compute than the expected complete-data log-likelihood
$Q(\theta |\theta ')$. Furthermore, whenever $\mathcal{R}I_1$ and
$\mathcal{R}I_0$ are
close to each other, as in our real-data examples, $\mathcal{R}I_{0.5}$
will be practically the same as either $\mathcal{R}I_1$ or $\mathcal{R}I_0$.

\subsection{An Illustrative Finite-Sample Comparison}\label{subsec:finite}

Let $Y_{\mathrm{co}}=\{y_1, \ldots, y_n\}$ be i.i.d. samples from
$N(\mu,\break \sigma^2)$, where both $\mu$ and $\sigma^2$ are unknown, and the
null hypothesis is $H_0\dvtx \mu= \mu_0$. Suppose our observed data
$Y_{\mathrm{ob}}$ is a size-$m$ random sample of $Y_{\mathrm{co}}$,
where \mbox{$0<m<n$}. Then it
should be clear that the relative information is $r=m/n$ by any
reasonable argument. Indeed, straightforward calculation shows
$\mathcal{R}I_1=r$ regardless of the actual value of
$Y_{\mathrm{ob}}$. However,
%
\begin{eqnarray}\label{normal}
\quad
\mathcal{R}I_0 &=& \frac{1}{r}
\biggl[1-\frac{\log (1+(1-r^2)({t_0^2}/{m}) )}
{\log (1+({t_0^2}/{m}) )}  \biggr]
\nonumber\\[-8pt]
\\[-8pt]
&=& r -\frac{r(1-r^2)}{2}\frac{t_0^2}{m}
+ O \biggl( \biggl(\frac{t^2_0}{m} \biggr)^2 \biggr),
\nonumber
\end{eqnarray}
where $t_0 = (\bar y_m - \mu_0)/\sqrt{\hat\sigma_m^2/m}$, which
differs from the usual $t$-statistic (under the null) only due to
the use of MLE for $\sigma^2$, $\hat\sigma^2_m= (1-1/m)s^2_m$,
instead of the sample variance $s^2_m$. From (\ref{normal}), it is
clear that $\mathcal{R}I_0$ approaches~$r$ whenever $t_0^2/m$ is small,
which implies that $\mathcal{R}I_0$ will recover (reasonably) the correct
information when the null hypothesis is (approximately) correct.

In contrast, for a fixed sample size $m$, $\mathcal{R}I_0$ approaches zero
if $t_0^2 \rightarrow\infty$ because for large $t_0^2$, $\mathcal{R}I_0$
behaves like $-r^{-1}\log(1-r^2)/\log(1+\frac{t^2_0}{m})$. The
reason is that the larger $t_0^2$ is, the stronger is the \mbox{evidence}
that the null is false, and thus the more conservative we become
when we impute $\mathrm{lod}(\mu, \mu_0|Y_{\mathrm{co}})$ using
$\mathrm{E} [\mathrm{lod}(\mu,
\mu_0|Y_{\mathrm{co}}) |Y_{\mathrm{ob}}, \mu_0 ]$. In other words,
whereas $\mathcal{R}I_0$ is a
good measure of how conservative the inference is, this example demonstrates
that measuring conservatism in general is not necessarily the same
as measuring the relative information. However, when the true
$\theta $ is in a reasonable neighborhood of $\theta _0$, $\mathcal{R}I_0$ can
be a valuable measure, especially because it is more robust to the
posited alternative model and thus can serve as a useful
diagnostic measure complementing $\mathcal{R}I_1$. We also note the
potentially different impacts of nuisance parameter on $\mathcal{R}I_0$
and $\mathcal{R}I_1$. When $\sigma^2$ is known, $\mathcal{R}I_0=\mathcal{R}I_1=r$.
However, whereas $\mathcal{R}I_1$ remains the same when $\sigma^2$ is
unknown, $\mathcal{R}I_0$ is greatly affected.

It is also informative to see how $\mathcal{B}I_0$ of (\ref{eq:bisg}) and
$\mathcal{B}I_s$ of (\ref{eq:combss}) compare in this simple problem. For
reasons discussed previously, we fix here the nuisance parameter
$\sigma^2$ at its MLE under the null,
$\tilde\sigma^2_{\mathrm{ob}}=\sum_{i=1}^m (y_i-\mu_0)^2/m$.
We therefore effectively have a
single-parameter $\mu$, whose score function given a normal sample
$\{y_1, \ldots, y_m\}$ is $S_m(\mu)=m(\bar y_m - \mu)/\sigma^2$
(where $\sigma^2$ is treated as known). Using the fact that
$I_{\mathrm{mi}}(\mu|Y_{\mathrm{ob}})=(n-m)/\sigma^2$, we have from
(\ref{eq:bisg}), after
setting $\sigma^2=\tilde\sigma^2_{\mathrm{ob}}$,
%
\begin{eqnarray}\label{eq:biot}
\quad
\mathcal{B}I_0 &=& \frac{ m^2(\bar y_m - \mu_0)^2/
\tilde\sigma^4_{\mathrm{ob}}}{m^2(\bar y_m - \mu_0)^2/
\tilde\sigma^4_{\mathrm{ob}}+ (n-m)/\tilde\sigma^2_{\mathrm{ob}}}
\nonumber\\[-8pt]
\\[-8pt]
&=& \frac{rt^2_0}{r t^2_0 + (1-r)(1+t^2_0/m)}.
\nonumber
\end{eqnarray}
It should not be a surprise to see that $\mathcal{B}I_0=0$ when $t_0=0$,
that is, when $\mu_0$ happens to be the MLE of $\theta $, $\bar y_m$, a
phenomenon we previously noted in Section~\ref{subsec:shrink}.
However, this simple example provides some clues on why this
happens.

Recall that $\mathcal{B}I_0$ was derived by assuming that the prior
shrinks to the null. This is very strong prior information, and it
inevitably influences our measure of the relative information.
Consider the situation when $t_0=0$, in which case our observed
data are completely consistent with our strong prior that
$\theta =\theta _0$. In that sense, the information from the observed data
is completely useless because it does not provide anything more
than we a priori knew (or rather, assumed). Hence it is not
a contradiction for $\mathcal{B}I_0$ to declare zero relative information
when clearly the relative information in the observed data should
be $r$. It is not a contradiction because $\mathcal{B}I_0$ has
incorporated the prior information, whereas $r=m/n$ measures the
relative information in the data under our posited model. This
\mbox{argument} appears to be further substantiated when we consider the
other extreme, namely, when $t^2_0 \rightarrow\infty$. By the
same logic, in this case, the observed data are extremely
informative as they provide strong evidence to contradict the
prior, and the degree of contradiction is such that, even with
more data, it is unlikely to be altered. Consequently, one can
expect $\mathcal{B}I_0$ to be close to $1$, which indeed follows from
(\ref{eq:biot}) when $m$ is large because $\mathcal{B}I_0 \rightarrow
[1 +(r^{-1}-1)m^{-1}]^{-1}$ when $t^2_0\rightarrow\infty$.

The above discussion indicates a potential problem with any
Bayesian measure, as it is inevitable that some prior information
will ``leak'' into our measure of relative information in the data
alone (for a specified test). When we have reliable prior
information, it is a very interesting issue to investigate/debate
whether our relative information should include the prior
information (e.g., in the extreme case when we know the null is
true for certain, the data become irrelevant, and one can always
consider we have 100\% information). Nevertheless, in cases where
the prior is introduced for convenience, as largely the case for
our setting, it is desirable to reduce any unintended influence as
much as possible. In this regard, it was a pleasant surprise to
see that the $\mathcal{B}I_s$ defined in (\ref{eq:combss}) is able to
recover the correct answer in this example. Specifically, letting
$\sigma^2=\tilde\sigma^2_{\mathrm{ob}}$, (\ref{eq:combss}) becomes
%
\begin{eqnarray}\label{eq:bios}
\qquad
\mathcal{B}I_s
&=& \frac{ \sum_{i=1}^m(y_i - \mu_0)^2/\tilde\sigma^4_{\mathrm{ob}}}
{\sum_{i=1}^m(y_i - \mu_0)^2/\tilde\sigma^4_{\mathrm{ob}}+ (n-m)/
\tilde\sigma^2_{\mathrm{ob}}}
\nonumber\\[-8pt]
\\[-8pt]
&=& \frac{m}{m +(n-m)}=r.
\nonumber
\end{eqnarray}
It is curious that $\mathcal{B}I_s$ has this ability of ``removing'' the
impact of prior information that affected $\mathcal{B}I_0$ in this
finite-sample setting; how generally this result holds (even
approximately) is a topic for future research.

\subsection{Connections to the Two CR Information Lower~Bounds}

Our large-sample measures have interesting connections with classic
measures based on Fisher information, as shown in Section~\ref{subsec:asym}.
Are there similar connections for the small-sample Bayesian measures?
The Bayesian measures are based on
posterior variances of likelihood ratios or their logarithms. It
turns out that there are several interesting connections, or at
least analogies, in both frequentist and Bayesian literature. In a
frequentist setting, just as the well-known Cram\'er--Rao lower bound
provides a finite-sample information bound that is determined by
the Fisher information, there is a more general Chapman--Robbins
information bound (\cite*{chap51}) that is based on
sampling variance of the \textit{likelihood ratio}. Specifically, let
$X$ have a multivariate pdf/pmf $f(X|\theta)$ with $\theta$
taking values in some parameter space~$\Theta$. For each $\theta$, let
$S_\theta=\{x\dvtx f(x|\theta)>0\}$ be the support of $f(X|\theta)$.
Suppose $T(X)$ is an unbiased estimator of a real-valued function
$\tau(\theta)$. Let
\[
\Phi_\theta =\{\phi\in\Theta\dvtx \tau(\phi)\not=\tau(\theta)
\mbox{ and } S_\phi\subset S_\theta\}.
\]
Then
\[
\operatorname{Var}(T(X)|\theta) \ge\sup_{\phi\in\Phi_\theta}
 \biggl[{[\tau(\phi) - \tau(\theta)]^2\over\operatorname{Var}
 (LR(\phi,\theta|X)|\theta) }\biggr],
\]
where $LR(\phi, \theta|X)$ denotes the likelihood ratio function
$f(X|  \phi)/f(X|\theta)$.

This ``second CR'' bound is more general than the first one
because it requires neither differentiability of $\tau(\theta)$
nor the existence of Fisher information (e.g., as in the case of
discrete parameters). It provides an interesting analogy to the
proposed Bayesian measures because it is based also on the
variability of the likelihood ratio, where $\phi$ and $\theta$ can
be arbitrarily apart. The central connection here is that while
our large-sample measures have close ties with Fisher information
(as detailed in Section \ref{subsec:asym}), which is also intimately connected
with the ``first CR'' bound (i.e., Cram\'er--Rao bound), our
small-sample measures are based on variances of likelihood ratio,
which is connected with the ``second CR'' bound. The fact that the
second CR bound is more general\vadjust{\goodbreak} than the first CR bound is also
consistent with our expectation that our Bayesian measures
ultimately should be more general than the likelihood-based
large-sample measures, though currently this is still just an
expectation, not a realization.

\subsection{Connections Between Likelihood Ratio and Bayes Factors}

The variances in our Bayesian measures are more general than the
one used by the second CR bound because we average over not only
the missing data but also the posterior distribution of $\theta $.
Examining the posterior distribution of the entire likelihood
ratio might seem a case of ``using data twice,'' but the
following several identities suggest that such a practice is
natural from the Bayesian point of view (indeed, the use of
posterior distribution of the likelihood ratio has been previously
advocated by \cite*{demp97}).

First, suppose we have a \emph{proper} prior $\pi(\theta)$;
then it is easy to verify that
%
\begin{eqnarray}\label{bafac}
&& \mathrm{E}[\mathrm{LR}(\theta _0,\theta |Y_{\mathrm{ob}})|Y_{\mathrm{ob}}]
\nonumber \\
&&\quad
= \int\frac{f(Y_{\mathrm{ob}}|\theta _0)}{f(Y_{\mathrm{ob}}|
\theta)}\frac{f(Y_{\mathrm{ob}}|\theta )\pi(\theta)}
{f_\pi(Y_{\mathrm{ob}})}\,d\theta
\\
&&\quad
= \frac{f(Y_{\mathrm{ob}}|\theta _0)}{f_\pi(Y_{\mathrm{ob}})}
\equiv\mathrm{BF}_{\mathrm{ob}},
\nonumber
\end{eqnarray}
where $f_\pi(Y_{\mathrm{ob}}) = \int f(Y_{\mathrm{ob}}|\theta )\pi
(\theta ) \,d\theta $. (Note that here
we assume $\theta _0$ is fixed at a known value.)

In other words, the posterior mean of our likelihood ratio is
simply the well-known Bayes factor for assessing the probability
of the model under $\theta =\theta _0$ relative to the model under
$\theta \sim \pi(\theta )$. This shows that the Bayes factor is a very natural
generalization of likelihood ratio by taking into account our
uncertainty in~$\theta $ while accessing the evidence in the data
against the hypothesized null value $\theta =\theta _0$. It also shows
that it is quite natural to consider posterior quantification of
the likelihood ratio itself. Incidentally, applying identity
(\ref{bafac}) first with $Y_{\mathrm{ob}}=Y_{\mathrm{co}}$ and then
averaging the
resulting identity over the posterior predictive distribution
$f(Y_{\mathrm{co}}|Y_{\mathrm{ob}})$, we also obtain the following
intriguing result:
%
\begin{eqnarray}\label{bayes}
\qquad
\mathrm{E}[\mathrm{BF}_{\mathrm{co}}|Y_{\mathrm{ob}}]
&=& \mathrm{E}[\mathrm{LR}(\theta _0, \theta
|Y_{\mathrm{co}})|Y_{\mathrm{ob}}]
\nonumber\\[-8pt]
\\[-8pt]
&=& \mathrm{E}[\mathrm{LR}(\theta _0, \theta |Y_{\mathrm{ob}})|
Y_{\mathrm{ob}}] =\mathrm{BF}_{\mathrm{ob}}.
\nonumber
\end{eqnarray}
In other words, the observed-data Bayes factor $\mathrm{BF}_{\mathrm{ob}}$ is the
posterior average of any of these three quantities: the
observed-data likelihood ratio, the complete-data likelihood
ratio, or the complete-data Bayes factor. Identities
(\ref{ratio}), (\ref{bafac}) and (\ref{bayes}) together
demonstrate the ``coherence'' of likelihood ratio and Bayes factor
as well as between them. Identity (\ref{bayes}) also suggests an
easy way of computing $\mathrm{BF}_{\mathrm{ob}}$ via Monte Carlo
averaging of
complete-data or observed-data likelihood ratios. We note,
however, that the\break posterior distributions of
$\mathrm{BF}_{\mathrm{co}}$, $\mathrm{LR}(\theta _0,\theta |Y_{\mathrm{co}})$
and $\mathrm{LR}(\theta _0, \theta|Y_{\mathrm{ob}})$
are generally different. In
particular, because of (\ref{ratio}) and (\ref{bafac}), we have
that
%
\begin{eqnarray}\label{eq:vari}
\quad
&& \max \{\operatorname{Var} [\mathrm{BF}_{\mathrm{co}}|
Y_{\mathrm{ob}} ], \operatorname{Var}[\mathrm{LR}(\theta _0,
\theta  |Y_{\mathrm{ob}})|Y_{\mathrm{ob}} ] \}
\nonumber\\[-8pt]
\\[-8pt]
&&\quad \leq \operatorname{Var} [\mathrm{LR}(\theta _0,
\theta  |Y_{\mathrm{co}})|Y_{\mathrm{ob}} ].
\nonumber
\end{eqnarray}

Given the clear interpretation and utility of the posterior mean
of the likelihood ratio, we would naturally consider the posterior
variance of the likelihood ratio. That is, we can measure the
posterior uncertainty in our likelihood ratio evidence. These are
exactly the quantities used in defining ${\mathcal B}I_1^\pi$ in
(\ref{ripi}), where the numerator and denominator are respectively the
posterior variances of the observed-data and complete-data
likelihood ratios. The following equivalent
expression of ${\mathcal B}I_1^\pi$ further demonstrates how
${\mathcal B}I_1^\pi$ measures relative ``flatness'' in the
likelihood ratio surfaces:
%
\begin{equation}\label{bicov}
\quad
{\mathcal B}I_1^\pi= \frac{\operatorname{Cov}_\pi
[\mathrm{LR}(\theta _0,\theta|Y_{\mathrm{ob}}),
\mathrm{LR}(\theta , \theta _0|Y_{\mathrm{ob}})]}
{\operatorname{Cov}_{\pi,\theta _0}
[\mathrm{LR}(\theta _0,\theta|Y_{\mathrm{co}}),
\mathrm{LR}(\theta , \theta _0|Y_{\mathrm{co}})]},
\hspace*{-20pt}
\end{equation}
where $\operatorname{Cov}_\pi$ is the covariance operator with
respect to the
prior $\pi(\theta )$, and $\operatorname{Cov}_{\pi, \theta _0}$ is
with respect to
$f(Y_{\mathrm{co}}|Y_{\mathrm{ob}}, \theta _0)\pi(\theta )$. In
other words, the flatness of the
likelihood ratio surfaces is measured by the covariance of the
likelihood ratio and its reciprocal. Although this expression
itself is intuitive because a positive function is flat if and
only if it is proportional to its reciprocal, the equivalence
between (\ref{ripi}) and (\ref{bicov}) is a bit curious because
(\ref{ripi}) is based on \textit{posterior variance} whereas
(\ref{bicov}) is based on \textit{prior covariance}.

\subsection{Connections to Entropy and $R^2$}

It would be a serious oversight if we do not emphasize the
connections of the information measures we discuss in this paper
to the vast literature on entropy. Indeed, essentially all
measures we presented have an entropy flavor, from the
large-sample ones based on Kullback--Leibler information to the
small-sample ones involving second-order entropy in the form of
$\int(\log p(\theta ))^2 p(\theta ) \,d\theta $
(see \cite*{zell03}). This is
very natural given that the entropy is a fundamental type of information
measure (e.g., \cite*{akai85}).
Indeed, much of the classic results on information measure in
optimal sequential designs, which our genetic\vadjust{\goodbreak} applications
resemble (i.e., as one needs to decide the next step given what
has been observed), are based on entropy-like quantities and their
generalizations. This includes both Kullback--Leibler information
and Chernoff information (\cite*{cher79}). A central difference
between that literature and our current proposals is that the
existing literature focuses on quantifying the \textit{absolute}
amount of information in an experiment/design,\break whereas our main
objective here is to quantify the \textit{relative} amount of information
compared to the absolute amount of information that we would have if
there were no missing data (e.g., known IBD sharing in linkage studies).
Furthermore, we investigate two sets
of relative information, depending on whether we can assume the
true parameter is in a neighborhood of the null or not. To the
best of our knowledge, our study is the first serious
investigation of the roles of null and alternative hypotheses in
measuring relative information.

Because our Bayesian measures $\mathcal{B}I^\pi _1$ and $\mathcal{B}I^\pi _2$ are defined
as ratios of variances, it is also important to emphasize their
connections to the regression $R^2$ and to other measures of
association/correlation such as the linkage disequilibrium measure
$r^2$ (e.g., \cite*{devris95}). These measures are related to
Fisher information and can also be used to estimate relative
information. The main differences are that ours are defined via
the \emph{posterior variability} of the \emph{whole \mbox{likelihood}
ratio or log-likelihood ratio}, instead of \emph{sampling
variances} of \emph{individual statistics or variables}. More
details on measures of association/correlation used to quantify
relative information can be found elsewhere (\cite*{niinfo06}).

\section{Limitations and Further Work}\label{sec:future}

\subsection{Further Theoretical and Methodological Work}

Clearly much remains to be done, especially for the small-sample
problems. With large samples, we believe the measures we proposed,
especially $\mathcal{R}I_1$, satisfy essentially all five criteria as
discussed in \mbox{Section}~\ref{subsec:conf}. For small samples, the various Bayesian
measures we proposed, while all satisfy the second criterion, have
pros and cons regarding the rest of the criteria. The most
pronounced problem, of course, is the choice of a general-purpose
``default prior.'' Here we emphasize that the desire for ``general
purpose'' is motivated by the observation that in many applications
the investigators need to compute the\vadjust{\goodbreak} information measures for
many data sets (e.g., different families or pedigrees and
different loci in linkage analysis; different tests for different
haplotype models in the association studies) under time
constraints. Therefore it is typically not feasible to construct
specific priors for each data set at hand, nor is it desirable
given that the purpose of hypothesis testing, in the genetic
applications we are interested in, has more of a screening nature. A
requirement for constructing problem-specific priors would be
typically viewed as too much of a burden to be practically
appealing. On the other hand, standard recipes for constructing
``default'' priors do not seem to be generally applicable either.
For example, the use of Jeffreys' prior is typically out of the
question because the calculation of the expected Fisher
information requires us to specify a reliable distribution over
the state space of $Y_{\mathrm{ob}}$ for arbitrary value of $\theta
$, which is
typically very hard, if not impossible, to do. Furthermore, the
properties of Jeffreys' prior are not clear when we try to avoid
the use of Fisher information in the first place.

Second, whereas $\mathcal{B}I_s$ provides a nice connection between
small-sample and large-sample measures in the neighborhood of
$\theta _0$, we currently do not have such a measure when the null is
far from the truth. This is of great theoretical and practical
concern, at least in the context of genetic studies, because the
regions where there is strong evidence against the null are
precisely the regions we try to identify. One possible strategy is
to start by estimating $\theta $ based on the aggregated data (e.g.,
using data from the other families), and then use a
prior that shrinks toward this estimated $\theta $ when computing
information measure for individual components (e.g., families).
In future work we plan to evaluate this strategy, as a part of
the general investigation of the sensitivity of our Bayesian measures
to prior
specifications once we move out the neighborhood of the null.

Third, even for large samples, our measures $\mathcal{R}I_0$ and
$\mathcal{R}I_1$
can be sensitive to the posited linkage or association model,
which may or may not capture the real biological process that
leads to the linkage or association. This would be particularly
true for $\mathcal{R}I_1$, which relies more heavily on the model
associated with the test than $\mathcal{R}I_0$. Although such sensitivity
is inevitable because without a specific alternative model the
very notion of relative information may not even be defined, as we
emphasized previously, it is important to understand to what
degree our information measures can change with our working model.
Both theoretical and empirical investigations are needed,
especially for classes of problems that are common in practice.
Also needed are investigations of the impact of nuisance
parameters on these measures. The haplotype association examples
involve nuisance parameters, for example, population genotype risks or
population haplotype frequencies, and $\mathcal{R}I_1$ seems to work
adequately in practice. Nevertheless, it would be interesting to
see if further refinements are possible. The illustrative example
of Section~\ref{subsec:finite} strongly suggests that further
research is necessary to investigate the possible complications
caused by the nuisance parameters, especially for $\mathcal{R}I_0$.

\subsection{Other Applications}

The genetic applications presented in this paper focus on the
allele-sharing linkage methods and the haplotype-based association
studies, but there are many other areas in genetics where measuring
relative information is important. For example, in the past
years the markers used in genome-wide searches for susceptibility
loci were mostly microsatellites. These are markers that have many
alleles, and are generally very informative, but are not
very common across the genome. Because the applications
focused on small regions of the genome, this lack of abundance of
the microsatellites has led to the still increasing popularity of
the SNPs as genetic markers. The SNPs are not as informative as
the microsatellites, but they are highly abundant. Also new
technology platforms such as the Affymetrix GeneChip Mapping 10K, 100K and
500K Arrays (\cite*{matsuzaki04}) are
available for SNP genotyping, and they come with a substantial
reduction in cost. Given that both the microsatellites and the
SNPs are currently used in gene-mapping studies, a fundamental
and practical question is how many SNPs we need in order to obtain
the same amount of information as obtained by using microsatellites.
Differences between SNPs and microsatellites have been
investigated for linkage
(e.g., \cite*{krug97}; \cite*{scha04}; \cite*{evans04};
\cite*{midd04}; \cite*{thala05}), and
measures of relative information
extracted have been proposed (\cite*{tengsieg98}), but the answers
to similar questions will be different for different applications.
We plan to further explore the use of the proposed
measures of information to other problems of this sort. The
comparisons between the relative information of sets of SNPs to
that of sets of microsatellites (relative to the underlying
complete information) will allow us to make sensible comparisons
of the maps for a particular study purpose.

The gene-mapping research has focused recently on genome-wide
association studies that are thought to have better power to localize genes
contributing more modestly to disease susceptibility. In these studies, new
measures are needed for quantifying the loss in information due to untyped
SNPs, or even SNPs that have not been discovered.
Also, novel tools for measuring information are necessary in choosing a
subset of
``tagging'' SNPs to type for a disease project based on the data
from the HAPMAP project (\cite*{hapmap}).

Other possible applications are in testing for gene-environment
interaction. This can be done in both linkage and association
studies, and can increase the power of detecting risk factors. In
most of these studies, the environmental and the clinical data are
also incomplete. A natural question then arises: ``what is the
most efficient way to allocate the resources: what percentage
should be devoted to collect more genetic information and what
percentage should be used to collect more covariate information?''
The answer depends again on the specific study, and the problem is
more complicated because the environmental and clinical
information can be subject to much more complicated missing-data
patterns, often due to unknown reasons. Research is clearly
needed in this direction to explore to what extent it is possible
to sensibly measure the relative information for guiding the
allocation of resources, and we hope the general framework we set
up in this paper provides a starting point, if not a solution.

\begin{appendix}
\def\theequation{\arabic{equation}}
\setcounter{subsection}{0}

\section*{Appendix}

\subsection{\texorpdfstring{Proof for Section
\protect\ref{subsec:shrink}}{Proof for Section 5.2}}
\label{sA.1}

In order to prove the shrinking prior limit results in
Section~\ref{subsec:shrink}, we need the following lemma.
\begin{lemma}\label{lemma:lhopital}
Let $t$ be a fixed real number, and let $a_i$ and $b_i$,
$i=1,2,3,4$, be real continuous functions defined on an open
interval containing $t$, such that $a_i$ and $b_i$ are three times
differentiable in a neighborhood of $t$. Let $\tilde
a_i(\delta;t)=\int_{t-\delta}^{t+\delta}a_i(x)\,dx$, and similarly
for $\tilde b_i(\delta;t)$, where $i=1,2,3,4$. If
\setcounter{equation}{46}
\begin{eqnarray}\label{eq:zero}
a_1(t)a_2(t) &=&a_3(t)a_4(t), \quad
\nonumber\\[-8pt]
\\[-8pt]
b_1(t)b_2(t)&=&b_3(t)b_4(t),
\nonumber
\end{eqnarray}
but
%
\begin{eqnarray}\label{eq:noze}
&& b_1''(t)b_2(t)+b_1(t)b_2''(t)
\nonumber\\[-8pt]
\\[-8pt]
&&\quad {} -b_3''(t)b_4(t)-b_3(t)b_4''(t)\neq0,
\nonumber
\end{eqnarray}
then
\begin{eqnarray*}
&& \lim_{\delta\rightarrow0}
\frac{\tilde a_1(\delta;t) \tilde a_2(\delta;t)- \tilde
a_3(\delta;t)\tilde a_4(\delta;t)}{\tilde b_1(\delta;t)
\tilde b_2(\delta;t)- \tilde b_3(\delta;t)\tilde b_4(\delta;t)}
\\
&&\quad =   (a_1''(t)a_2(t)+a_1(t)a_2''(t)
\\
&&\hspace*{24pt}{} -a_3''(t)a_4(t)
-a_3(t)a_4''(t))
\\
&&\qquad {}\cdot (b_1''(t)b_2(t)+b_1(t)b_2''(t)
\\
&&\hspace*{34pt}{}
- b_3''(t)b_4(t)-b_3(t)b_4''(t))^{-1}.
\end{eqnarray*}
\end{lemma}
\begin{pf}
The proof follows from the simple Taylor expansion
\[
\tilde a_i(\delta; t) = 2a_i(t)\delta+
\tfrac{1}{3}a_i''(t)\delta^3 + O(\delta^5),
\]
and conditions (\ref{eq:zero}) and (\ref{eq:noze}).
\end{pf}
\begin{prop}\label{prop:bi1}
Let $\pi$ be $U(\theta _0-\delta, \theta _0 + \delta)$. Then
%
\begin{eqnarray}\label{eq:blim}
\lim_{\delta\rightarrow0} \mathcal{B}I^\pi _k
&=& \frac{S^2(\theta_0|Y_{\mathrm{ob}}) }
{S^2(\theta _0|Y_{\mathrm{ob}}) + I_{\mathrm{mi}}(\theta_0|Y_{\mathrm{ob}})},
\nonumber\\[-8pt]
\\[-8pt]
\eqntext{\quad k=1,2.}
\end{eqnarray}
\end{prop}
\begin{pf}
Let $a_1(\theta )\equiv b_1(\theta ) = \exp [
\mathrm{lod}(\theta ,\theta _0|Y_{\mathrm{ob}}) ]$,
$b_2(\theta )=\mathrm{E}[\exp [ \mathrm{lod}
(\theta _0,\theta |Y_{\mathrm{co}}) ]|Y_{\mathrm{ob}},\theta _0]$
and $a_2(\theta )=a_1^{-1}(\theta )$. Then, as in (\ref{bicov}), it is
straightforward to verify that
%
\begin{equation}\label{eq:var-lr-obs}
\qquad
\mathcal{B}I^\pi _1 = \frac{\int a_1(\theta )\pi(\theta )\,d\theta
\int a_2(\theta )\pi(\theta )\,d\theta - 1}
{\int b_1(\theta )\pi(\theta )\,d\theta \int b_2(\theta)
\pi(\theta ) \,d\theta -1}.
\end{equation}
We can then apply Lemma \ref{lemma:lhopital} with
$a_3=a_4=b_3=b_4\equiv1$. The result for $k=1$ in (\ref{eq:blim})
then follows because
\begin{eqnarray*}
a_1''(\theta _0)&=&\ell''(\theta _0|Y_{\mathrm{ob}})+S^2(\theta_0|Y_{\mathrm{ob}}),
\quad
\\
a_2''(\theta _0)&=&-\ell''(\theta _0|Y_{\mathrm{ob}})+S^2(\theta_0|Y_{\mathrm{ob}})
\end{eqnarray*}
and
\begin{eqnarray*}
b_2''(\theta _0)
&=& \mathrm{E} [ -\ell''(\theta _0|Y_{\mathrm{co}})
+ \ell'^2(\theta _0|Y_{\mathrm{co}}) |Y_{\mathrm{ob}},\theta _0 ]
\\
&=& 2 I_{\mathrm{mi}}(\theta _0|Y_{\mathrm{ob}})
- \ell''(\theta_0|Y_{\mathrm{ob}}) + S^2(\theta _0|Y_{\mathrm{ob}}).
\end{eqnarray*}
Note that condition (\ref{eq:zero}) holds because
$a_i(\theta _0)=\break  b_i(\theta _0)=1$ for all $i$.

For $k=2$, the limit can be calculated by observing that
\begin{eqnarray*}
\mathcal{B}I^\pi _2 &=& \biggl(1+\operatorname{Var}
\biggl[\log\frac{P(Y_{\mathrm{co}}|Y_{\mathrm{ob}},\theta )}
{P(Y_{\mathrm{co}}|Y_{\mathrm{ob}},\theta _0)}\bigg|Y_{\mathrm{ob}} \biggr]
\\
&&\hspace*{40pt}\Big/
\operatorname{Var} [\mathrm{lod}(\theta , \theta _0
|Y_{\mathrm{ob}})|Y_{\mathrm{ob}} ]\biggr)^{-1}
\end{eqnarray*}
and then
calculating the limit of the ratio in the denominator. A little
algebra shows that this ratio can be expressed as
%
\begin{eqnarray}\label{eq:brat}
\qquad
&& \biggl(\int a_1(\theta )\pi(\theta )\,d\theta \int a_2(\theta )
\pi(\theta )\,d\theta
\nonumber \\
&&\hspace*{53pt}{} - \biggl[\int a_3(\theta )\pi(\theta )\,d\theta \biggr]^2\biggr)
\nonumber\\[-8pt]
\\[-8pt]
&&\quad {}\cdot
\biggl(\int b_1(\theta )\pi(\theta )\,d\theta \int b_2(\theta )\pi(\theta )
\,d\theta
\nonumber \\
&&\quad \hspace*{61pt}{}
- \biggl[\int b_3(\theta )\pi(\theta )\,d\theta \biggr]^2\biggr)^{-1},
\nonumber
\end{eqnarray}
where $a_1(\theta )=b_1(\theta )$ are the same as in
(\ref{eq:var-lr-obs}), but
\begin{eqnarray*}
a_2(\theta )&=& \mathrm{E} \bigl[ \bigl(\mathrm{lod}(\theta ,\theta
_0|Y_{\mathrm{co}})-\mathrm{lod}(\theta ,\theta _0|Y_{\mathrm{ob}})\bigr)^2
\\
&&\hspace*{30pt}{}\cdot
\exp(\mathrm{lod}(\theta ,\theta _0|Y_{\mathrm{co}}))
|Y_{\mathrm{ob}}, \theta _0 \bigr],
\\
a_3(\theta ) &=& \mathrm{E} \bigl[\bigl(\mathrm{lod}(\theta ,\theta
_0|Y_{\mathrm{co}})-\mathrm{lod}(\theta ,\theta _0|Y_{\mathrm{ob}})
\bigr)
\\
&&\hspace*{28pt}{}\cdot
\exp(\mathrm{lod}(\theta ,\theta _0|Y_{\mathrm{co}}))
 |Y_{\mathrm{ob}},\theta _0 \bigr],
\\
b_2(\theta ) &=& \mathrm{lod}^2(\theta ,\theta _0|
Y_{\mathrm{ob}})a_1(\theta )
\quad\hbox{and}\quad
\\
b_3(\theta )&=& \mathrm{lod}(\theta ,\theta _0|
Y_{\mathrm{ob}})a_1(\theta ).
\end{eqnarray*}
To apply Lemma~\ref{lemma:lhopital}, we let
$a_4=a_3$ and $b_4=b_3$. Noting that $a_i(\theta _0)=b_i(\theta_0)=0$ for
all $i=2,3,4$ [and hence condition (\ref{eq:zero}) holds], we only
need to compute $a_2''(\theta _0)$ and $b_2''(\theta _0)$ in order to
obtain the limit. This calculation is facilitated by the formula
\begin{eqnarray*}
&& \frac{d^2}{dx^2} [ g^2(x)\exp(f(x))]
\\
&&\quad =    2g'^2\exp(f)+2gg''\exp(f)+4gg'f'\exp(f)
\\
&&\qquad {} +g^2f''\exp (f)+g^2f'^2\exp(f).
\end{eqnarray*}
The result then follows because
\[
b_2''(\theta _0)=2 \ell'^2(\theta _0|Y_{\mathrm{ob}})
= 2 S^2(\theta_0|Y_{\mathrm{ob}})
\]
and
\begin{eqnarray*}
a_2''(\theta _0)&=& 2 \mathrm{E}
\bigl[\bigl(S(\theta _0|Y_{\mathrm{co}})-S(\theta _0|Y_{\mathrm{ob}})\bigr)^2
 |Y_{\mathrm{ob}},\theta _0\bigr]
\\
&\equiv& 2I_{\mathrm{mi}}
(\theta_0|Y_{\mathrm{ob}}).
\end{eqnarray*}\upqed
\end{pf}

\subsection{\texorpdfstring{Derivations for Section
\protect\ref{subsec:fixed-size}}{Derivations for Section 6.2}}
\label{sA.2}

The derivations are based on the following lemma, which is trivial
to verify using the Taylor expansion.
\begin{lemma}\label{lemma:epsilons}
Let $f$ and $g$ be continuous functions defined on an open
interval containing zero, such that
$f(\delta)=a_1+a_2\delta+O(\delta^2)$ and
$g(\delta)=b_1+b_2\delta+O(\delta^2)$ as $\delta\rightarrow0$.
Then
\[
\frac{f(\delta)}{g(\delta)}=\frac{a_1}{b_1}+\frac
{a_2-b_2(a_1/b_1)}{b_1}\delta +O(\delta^2).
\]
\end{lemma}

As in Section \ref{sec:small}, we let $\delta=\theta _0 - \theta _{\mathrm{ob}}$.
For $\mathcal{R}I_1$, we
need to expand both $\ell_{\mathrm{ob}}(\theta _0)$ and
$Q(\theta_0|\theta _{\mathrm{ob}})$, as
functions of $\delta$. Using the notation given in
Section~\ref{subsec:fixed-size} and~(\ref{eq:emid}), we have
%
\begin{equation}\label{eq:lodex}
\qquad
\ell_{\mathrm{ob}}(\theta _0) -\ell_{\mathrm{ob}}(\theta_{\mathrm{ob}})
= - {I_{\mathrm{ob}}\over2} \delta^2 +
{\ell^{(3)}_{\mathrm{ob}}\over6} \delta^3 + O(\delta^4)
\hspace*{-9pt}
\end{equation}
and
%
\begin{eqnarray}\label{eq:qex}
&& Q(\theta _0|\theta _{\mathrm{ob}})
- Q(\theta _{\mathrm{ob}}|\theta _{\mathrm{ob}})
\nonumber\\[-8pt]
\\[-8pt]
&&\quad = - {I_{\mathrm{co}}\over2} \delta^2
+ {Q^{(3,0)}_{\mathrm{ob}}\over6} \delta^3 + O(\delta^4).
\nonumber
\end{eqnarray}
Expansion (\ref{eq:rione}) then follows directly from
Lemma~\ref{lemma:epsilons}.

To establish a similar expansion for $\mathcal{R}I_0$, let
$\theta_Q$ be the maximizer of $Q(\theta |\theta _0)$;
recall we assume that $\theta_Q$ is unique. Then
%
\begin{equation}\label{fracmc}
\mathcal{R}I_0= \frac{Q(\theta_Q|\theta _0)
- Q(\theta _0|\theta _0)}{\ell_{\mathrm{ob}}
(\theta _{\mathrm{ob}})-\ell_{\mathrm{ob}}(\theta _0)}.
\end{equation}
However, even when $\delta= \theta _0 - \theta _{\mathrm{ob}}$ is
small, it is not
immediate that $\theta _Q$ would be close to $\theta _{\mathrm{ob}}$
as well. We now
show that when $\delta$ is small enough,
$Q^{(1,0)}(\theta _0|\theta _0)$ and
$Q^{(1,0)}(\theta _{\mathrm{ob}}|\theta _0)$ have
opposite signs. Consequently, $\theta _Q$, the unique solution of
$Q^{(1,0)}(\theta |\theta _0)=0$, must be between $\theta _0$ and
$\theta _{\mathrm{ob}}$, and hence $|\theta _Q-\theta _{\mathrm{ob}}|\le|\delta|$.

To see this, we first expand $g(\theta )\equiv Q^{(1,0)}(\theta
|\theta )$
around $g(\theta _{\mathrm{ob}})$ to obtain
%
\begin{eqnarray}\label{eq:apB-1}
\qquad
g(\theta _0)-g(\theta _{\mathrm{ob}})
&=& g^{(1)}(\theta _{\mathrm{ob}}) \delta+ O(\delta^2)
\nonumber\\[-8pt]
\\[-8pt]
&=& \bigl[Q^{(2,0)}_{\mathrm{ob}}+ Q^{(1,1)}_{\mathrm{ob}}\bigr]
\delta+ O(\delta^2).
\nonumber
\end{eqnarray}
But the following general result, proved in \citet{meng00b}:
%
\begin{eqnarray}\label{eq:meng}
\ell^{(k+1)}_{\mathrm{ob}}(\theta )
&=& \sum_{j=0}^k  \pmatrix{%
j \cr k }Q^{(j+1,k-j)}(\theta |\theta )
\nonumber\\[-8pt]
\\[-8pt]
\eqntext{\quad\hbox{for any } k\ge 0,}
\end{eqnarray}
implies that $g(\theta _{\mathrm{ob}})=0$ and $Q^{(2,0)}_{\mathrm{ob}}+
Q^{(1,1)}_{\mathrm{ob}}= \ell^{(2)}_{\mathrm{ob}}= -I_{\mathrm{ob}}$. Consequently,
%
\begin{equation}\label{eq:apB}
Q^{(1,0)}(\theta _0|\theta _0) = -I_{\mathrm{ob}} \delta+ O(\delta^2).
\end{equation}
For $Q^{(1,0)}(\theta _{\mathrm{ob}}|\theta _0)$, using the notation in
(\ref{eq:emex}) and (\ref{eq:dq}), we have
%
\begin{eqnarray}\label{eq:apB-2}
&& Q^{(1,0)}(\theta _{\mathrm{ob}}|\theta _0)
\nonumber \\
&&\quad = \ell^{(1)}_{\mathrm{ob}}(\theta _{\mathrm{ob}})
+ H^{(1,0)}(\theta _{\mathrm{ob}}|\theta _0)
\nonumber\\[-8pt]
\\[-8pt]
&&\quad = H^{(2,0)}(\theta _0|\theta _0)(\theta _{\mathrm{ob}}
-\theta_0) + O(\delta^2)
\nonumber \\
&&\quad = I_{\mathrm{mi}}(\theta _0) \delta+O(\delta^2),
\nonumber
\end{eqnarray}
where $I_{\mathrm{mi}}(\theta )$ is as defined in (\ref{eq:imis}).
Since both
$I_{\mathrm{ob}}$ and $I_{\mathrm{mi}}(\theta _0)$ are positive, we
conclude from (\ref{eq:apB}) and~(\ref{eq:apB-2}) that
$Q^{(1,0)}(\theta _{\mathrm{ob}}|\theta _0)$
and $Q^{(1,0)}(\theta _0|\theta _0)$ have opposite signs when $\delta$
is small enough. Therefore we have established that
$\theta _Q-\theta _0=O(\delta)$, and consequently we can express
%
\begin{equation}\label{eq:delta}
\theta _Q -\theta _0 = B\delta+C\delta^2 + O(\delta^3),
\end{equation}
where $B$ and $C$ are $O(1)$ as
$\delta\rightarrow0$ and are to be determined.

To determine $B$ and $C$, we first note that
%
\begin{eqnarray}\label{eq:ellf}
Q^{(1,0)}(\theta _0|\theta _0)
&=& \ell_{\mathrm{ob}}^{(1)}(\theta_0)
\nonumber\\[-8pt]
\\[-8pt]
&=& - {I_{\mathrm{ob}}} \delta
+ {\ell^{(3)}_{\mathrm{ob}}\over2} \delta^2 + O(\delta^3)
\nonumber
\end{eqnarray}
and
%
\begin{eqnarray}\label{eq:qzero}
0 &=&Q^{(1,0)}(\theta _Q|\theta _0)
\nonumber\\
&=& \ell_{\mathrm{ob}}^{(1)}(\theta _0)+G^{(2)}(\theta _0)(\theta_Q-\theta _0)
\\
&&{} + {G^{(3)}(\theta _0)\over2}(\theta _Q-\theta _0)^2 + O(\delta^3),
\nonumber
\end{eqnarray}
where $G^{(k)}(\theta )\equiv Q^{(k,0)}(\theta |\theta )$. Substituting
(\ref{eq:delta}) and (\ref{eq:ellf}) into
(\ref{eq:qzero}) and solving for $B$ and $C$, we obtain
%
\begin{eqnarray}\label{eq:bc}
B&=& \frac{I_{\mathrm{ob}}}{G^{(2)}(\theta _0)}
\quad\hbox{and} \quad
\nonumber\\[-8pt]
\\[-8pt]
C &=& - \frac{\ell^{(3)}_{\mathrm{ob}}+ B^2
G^{(3)}(\theta _0)}{2G^{(2)}(\theta _0)}.
\nonumber
\end{eqnarray}

Noting that $G^{(1)}(\theta _0)=\ell^{(1)}(\theta _0)$ and
(\ref{eq:ellf}), we then obtain
\begin{eqnarray*}
&& Q(\theta _Q|\theta _0)-Q(\theta _0|\theta _0)
\\
&&\quad = G^{(1)}(\theta _0)(\theta _Q-\theta _0)
+ \frac{G^{(2)}(\theta _0)}{2} (\theta _Q-\theta _0)^2
\\
&&\qquad {} + \frac{G^{(3)}(\theta _0)}{6}(\theta _Q-\theta _0)^3 + O(\delta^4)
\\
&&\quad =
\biggl[-I_{\mathrm{ob}}B+ \frac{1}{2}B^2G^{(2)}(\theta _0) \biggr] \delta^2
\\
&& \qquad {}
+ \biggl[\frac{1}{2}B\ell^{(3)}_{\mathrm{ob}}- CI_{\mathrm{ob}}
\\
&&\hspace*{41pt}{} + BCG^{(2)}(\theta _0)+
\frac{1}{6}B^3G^{(3)}(\theta _0) \biggr] \delta^3 + O(\delta^4)
\\
&&\quad = - \frac{I^2_{\mathrm{ob}}}{2G^{(2)}(\theta _0)}\delta^2
+ \biggl[\frac{\ell_{\mathrm{ob}}^{(3)}I_{\mathrm{ob}}}{2G^{(2)}(\theta _0)}
+ \frac{G^{(3)}(\theta _0)I^3_{\mathrm{ob}}}{6 [G^{(2)}(\theta _0)]^3}
\biggr] \delta^3
\\
&&\qquad {}+ O(\delta^4).
\end{eqnarray*}
Combining this expansion with
\begin{eqnarray*}
G^{(2)}(\theta _0) &=& - I_{\mathrm{co}}
+ \bigl[Q^{(3,0)}_{\mathrm{ob}}+Q^{(2,1)}_{\mathrm{ob}}\bigr] \delta
+ O(\delta^2),
\\
G^{(3)}(\theta _0) &=& Q^{(3,0)}_{\mathrm{ob}}
+ \bigl[Q^{(4,0)}_{\mathrm{ob}}+Q^{(3,1)}_{\mathrm{ob}}\bigr] \delta
+ O(\delta^2)
\end{eqnarray*}
and applying Lemma~\ref{lemma:epsilons}, we obtain
\begin{eqnarray*} 
&& Q(\theta _Q|\theta _0)-Q(\theta _0|\theta _0)
\nonumber \\
&&\quad = \frac{I_{\mathrm{ob}}}{2}\mathcal{R}I_E \delta^2
\\
&&\qquad {}
+ \frac{1}{2}\biggl[\mathcal{R}I_E
\bigl[I_{\mathrm{ob}} \bigl(Q^{(3,0)}_{\mathrm{ob}}+Q^{(2,1)}_{\mathrm{ob}}
\bigr)-\ell^{(3)}_{\mathrm{ob}} \bigr]
\\
&&\qquad\hspace*{99pt} {}
- \frac{Q^{(3,0)}_{\mathrm{ob}}}{3}(\mathcal{R}I_E)^3 \biggr]
\delta ^3
\\
&&\qquad {} +O(\delta^4).
\end{eqnarray*}
By Lemma~\ref{lemma:epsilons}, the above equation and
(\ref{eq:lodex}) together imply that $\mathcal{R}I_0$ of (\ref{fracmc})
has the expansion (\ref{eq:rizer}).

\end{appendix}

\section*{Acknowledgments}

We thank Daniel Gudbjartsson for many helpful discussions and
suggestions, and Judy H. Cho for providing the inflammatory bowel
disease data. For the diabetes example illustrated in Figure \ref{fig:stroke}, we
thank Daniel Gudbjartsson for providing the software that
performed likelihood and information calculations, Gubmar
Thorleifsson for constructing the figure, and the diabetes
research group at Decode Genetics for generating and providing the
data. We also want to thank a number of reviewers for very
constructive comments and suggestions. This research was supported
in part by several National Science Foundation grants (Nicolae and
Meng).

\end{document}